\documentclass[pdflatex,sn-mathphys-num]{sn-jnl}

\usepackage[T2A,T1]{fontenc}
\usepackage[russian,english]{babel}

\usepackage{amsmath,amssymb,amsfonts,amsthm}
\usepackage{bbm}
\usepackage{xparse}

\usepackage{calligra}
\DeclareMathAlphabet{\mathcalligra}{T1}{calligra}{m}{n}
\DeclareFontShape{T1}{calligra}{m}{n}{<->s*[2.2]callig15}{}
\newcommand{\scriptr}{\mathcalligra{r}}

\usepackage{graphicx}
\usepackage{subcaption}
\usepackage{tikz}
\usetikzlibrary{calc,arrows.meta}
\tikzset{
  wbd/.style={line width=0.9pt},
  dot/.style={circle,fill=black,inner sep=1.4pt}
}
\usepackage{quantikz}

\IfFileExists{Orcidlogo.eps}{}{%
  \renewcommand{\orcidlogo}{\raisebox{0.15ex}{\textsf{\scriptsize ORCID}}}%
}

\usepackage{tabularx}
\usepackage{booktabs}
\usepackage[title]{appendix}
\usepackage{algorithm}
\usepackage{algorithmicx}
\usepackage{algpseudocode}
\usepackage{float}

\tikzset{
  wbd/.style={
    line width=0.9pt
  },
  dot/.style={
    circle,
    fill=black,
    inner sep=1.4pt
  },
  waveguide/.style={
    line width=1.1pt
  },
  resonator/.style={
    line width=1.2pt
  },
  quantum/.style={
    line width=1.1pt,
    -{Latex[length=2.5mm,width=1.8mm]}
  },
  tracebox/.style={
    draw,
    line width=1pt,
    rounded corners=2pt,
    minimum width=1.5cm,
    minimum height=1.1cm,
    inner sep=4pt
  }
}

\theoremstyle{thmstyleone}
\newtheorem{theorem}{Theorem}[section]
\newtheorem{lemma}[theorem]{Lemma}
\newtheorem{corollary}[theorem]{Corollary}

\theoremstyle{thmstyletwo}
\newtheorem{example}[theorem]{Example}
\newtheorem{remark}[theorem]{Remark}

\theoremstyle{thmstylethree}
\newtheorem{definition}[theorem]{Definition}

\NewDocumentEnvironment{questionbox}{o}
  {\begin{quote}\noindent\textbf{Question.}\ }
  {\end{quote}}

\begin{document}

\title[Minimal unitary representations of time-invariant causal quantum channels]
      {Minimal unitary representations of time-invariant causal quantum channels}

\author*[1]{\fnm{A. J.} \sur{Hasenack}\,
  \orcid{https://orcid.org/0009-0007-2304-6828}}
\email{a.j.hasenack@tue.nl}

\author[1]{\fnm{B.} \sur{{\v{S}}kori{\'c}}\,
  \orcid{https://orcid.org/0000-0003-1409-4127}}
\email{b.skoric@tue.nl}

\affil*[1]{
  \orgdiv{Department of Mathematics and Computer Science},
  \orgname{Eindhoven University of Technology},
  \orgaddress{
    \city{Eindhoven},
    \country{The Netherlands}
  }
}

\abstract{In this work, we study quantum channels that respect causality in single-input-single-output linear time-invariant systems. We show that positive semidefinite kernels satisfying certain analyticity constraints, which we call Schur-class kernels, provide a suitable description of such channels through multiplicative maps (Schur multipliers). Using the Moore-Aronszajn Theorem, we establish that these multipliers admit a minimal analytic Stinespring representation, which, in the unital case, is given by multiplication by a left inner function. We then relate the output rank of pure-state inputs to the dimension of the associated reproducing kernel Hilbert space. Using de Branges-Rovnyak realizations, we derive a stability criterion for the existence of two-sided inner completions of these vector-valued functions. We further show that the adjoint of the Julia embedding associated with the realization is minimal, and that its characteristic function essentially coincides with the canonical two-sided inner completion suggested by the Beurling-Lax-Halmos Theorem. Finally, we provide algorithms for explicitly computing minimal two-sided inner completions in the case of finite McMillan degree.}



\maketitle

\section{Introduction}

The idea that a signal may not exit a system before it enters, referred to as causality, is a fundamental constraint for modeling realistic physical systems. For a continuous-time input signal, $\psi$, and a linear time-invariant system, such constraints can be imposed in the frequency domain through analyticity of the corresponding transfer function, $h$, on the upper half-plane,
\[ (H \psi)(\omega) = h(\omega) \psi(\omega) \qquad h \text{ analytic on } \mathbb{C}^+\]
which acts pointwise on the input signal. For lossless quantum systems, such transformations should generally be unitary, which is equivalent to $|h(\omega)| = 1$ almost everywhere on $\mathbb{R}$.

It is clear that classical mixtures of such unitary transformations, $\varphi(\omega, \omega') = \sum_i p_i \overline{h_i(\omega)} h_i(\omega')$ with $p_i>0$ -- just like $h(\omega)$ -- act by multiplication on mixed states $\rho(\omega, \omega') \mapsto \varphi(\omega, \omega') \rho(\omega, \omega')$. The resulting map, $\varphi(\omega, \omega')$, is a positive semidefinite kernel, analytic in one variable and anti-analytic in the other (sesquianalytic). The above classical mixture motivates the general study of multiplicative maps, $M_\varphi$ (so-called Schur multipliers, cf. section \ref{sec:4.1}), where $\varphi$ is positive semidefinite, bounded and sesquianalytic on the upper half-plane. These kernels will henceforth be called \textit{Schur-class kernels} and they are the naturally arising entities when considering statistical mixtures of contractive causal systems (see Theorem \ref{thm:1}). Such multipliers, $M_\varphi$, are moreover related to the \textit{analytic Schur multipliers} in \cite{Aleksandrov2025}, while we impose sesquianalyticity instead of joint-analyticity, demand slightly weaker limiting behavior and additionally require positive semidefiniteness. 

An immediate question that arises is whether the definition of a Schur-class kernel is restrictive enough to yield a complete characterisation:
\begin{questionbox}
    Can every Schur-class kernel be implemented by a single causal transformation on an enlarged system?
\end{questionbox}
In Theorem \ref{thm:Fact_of_kernels} we establish that this is indeed the case, using a Kolmogorov factorization $ \varphi(\omega, \omega') = \langle v(\omega) | v(\omega') \rangle$ for some analytic vector-valued function $v(\omega)$ living in an enlarged Hilbert space. Such a factorization is moreover useful to prove the complete positivity of these maps, which is related to the \textit{measurable} Schur multiplier framework of Birman and Solomyak \cite{Birman2003}. It also allows us to regard $M_\varphi$ in the Heisenberg picture -- i.e. as a map on bounded linear operators -- and discuss its relation with the Stinespring representation Theorem. As it turns out, the naturally constructed $v$ induces a minimal Stinespring representation (corollary \ref{cor:Stinespring_rep}) and can be applied to study the spectral decomposition of output states for instance (corollary \ref{cor:spec_decomp}).

A next objective is of particular importance for quantum states: enforcing the trace preserving property (or unitality in the Heisenberg picture) onto the completely positive maps. To be precise, by \textit{time-invariant causal (TIC) quantum channel} we mean a unital Schur-class kernel as a multiplicative map. In line with our earlier question, we wish to answer
\begin{questionbox}
    When can a TIC quantum channel be implemented by a single causal, unitary transformation on an enlarged system?
\end{questionbox}
We resolve this question in two steps. First, we show that unitality forces the analytic Stinespring operator to be given by multiplication by a left inner function (Corollary \ref{cor:unital_real_schur_mult}). The question of unitary implementability therefore reduces to determining when this function admits a two-sided inner completion.

The Beurling-Lax-Halmos Theorem suggests a canonical candidate for such a completion. We establish a connection between its associated model space and that of the reflected function (Theorem \ref{thm:eq_of_model_spaces}). Using de Branges-Rovnyak realization theory, we further show that the adjoint of the corresponding Julia embedding is minimal and that its transfer function essentially coincides with the proposed completion (Theorem \ref{thm:eq_of_real}).

These results allow us to characterize two-sided inner completability through a stability criterion for the associated realization (Theorem \ref{thm:completability_and_main_operator}). We subsequently translate this criterion into a condition on the original Schur-class kernel, yielding a direct test of completability (Corollary \ref{cor:completability}).

The resulting framework is intended for the analysis of continuous-variable random scattering models arising in quantum circuits, communication and cryptography. For this reason, we put specific emphasis on the computational nature of our results. For the finite dimensional, rational case we provide algorithms that allow calculation of the factorization and two-sided inner completion and give an explicit example in section \ref{sec:application} as an application.

\section{Preliminaries of Quantum Information Theory}

Before we proceed, we provide some concepts of continuous-variable quantum information theory that we will require in our analysis. Such analysis differs from finite dimensional QIT that can be found in for example \cite{Watrous2018} or \cite{Nielsen2010}. We recommend \cite{Reed1980} for the general framework of (bounded) linear operators on Hilbert spaces and \cite{Holevo2012}, \cite{Wolf2023} and \cite{Paulsen2003} for extensions to QIT.

A central object in quantum information theory is the complex Hilbert space (a complete complex inner product space) which will often be denoted by $\mathcal{H}$ and the inner product by $\langle \cdot | \cdot \rangle$. In line with the Dirac notation, our inner product will be anti-linear in the first argument and linear in the second, unless specified otherwise. Moreover, it will always be assumed that the Hilbert space is separable, that is, it admits a countable, orthonormal basis. Typical examples are the space of square summable sequences, $\ell^2$, and square integrable functions over the field of real numbers, $L^2(\mathbb{R})$.

We denote the space of bounded, linear operators from one Hilbert space, $\mathcal{H}_A$, to another, $\mathcal{H}_B$, by $\mathcal{B}(\mathcal{H}_A, \mathcal{H}_B)$. For any $A$ in $\mathcal{B}(\mathcal{H}_A, \mathcal{H}_B)$, the adjoint is defined as the unique operator $A^\ast$ that satisfies $\langle A \psi | \phi \rangle_B = \langle \psi | A^\ast \phi \rangle_A$ for any $\psi \in \mathcal{H}_A$ and $\phi \in \mathcal{H}_B$\footnote{In fact, the space $\mathcal{B}(\mathcal{H})$ together with the adjoint (an involution) forms a structure called a $C^\ast$-algebra.}. The self-adjoint operators on $\mathcal{H}$ will be denoted by $\text{H}(\mathcal{H})$. An isometry from $\mathcal{H}_A$ to $\mathcal{H}_B$ is an operator, $U$ in $\mathcal{B}(\mathcal{H}_A, \mathcal{H}_B)$, with the property that its adjoint is the left inverse of $U$, i.e. $U^\ast U = I$. Similarly, a coisometry is defined such that its adjoint is the right inverse, $U U^\ast = I$. A partial isometry is a map that is isometric on the orthogonal complement of its kernel. $U$ is called unitary if it is both an isometry and a coisometry. The notion of compactness of an operator $K$ in $\mathcal{B}(\mathcal{H})$ can be characterized by the requirement that bounded sets, $B \subset \mathcal{H}$ are mapped to relatively compact sets (i.e. $\operatorname{clos}K(B)$ is compact). The space of compact operators on $\mathcal{H}$ is denoted by $\mathcal{K}(\mathcal{H})$ and is a Banach space. In particular, when $A$ is both a compact and normal operator (in the spectral sense), then $A$ has a countable set of eigenvectors.

An important subset of $\text{H}(\mathcal{H})$ is $\text{PSD}(\mathcal{H})$, the set of positive semidefinite operators, which satisfy that for any vector $\psi$ in $\mathcal{H}$ we have $\langle \psi | A \psi \rangle \geq 0$ and we write $A \geq 0$. For any set $S \subseteq \mathcal{B}(\mathcal{H})$, we will denote the subset of positive semidefinite elements by $S_+ := S \cap \text{PSD}(\mathcal{H})$. The trace of a positive semidefinite operator, $A$, is defined as $\operatorname{tr}[A] = \sum_{i=0}^{\infty} \langle e_i | A e_i \rangle$ and is independent of the chosen o.n.b., $\{e_i\}$, of $\mathcal{H}$. The just-defined trace induces two natural subsets (or to be precise, $\ast$-ideals) of $\mathcal{B}(\mathcal{H})$, the first one of which is called the Banach space of trace-class operators, $\mathcal{T}(\mathcal{H})$, for which each operator, $A$, is such that $\operatorname{tr}[\sqrt{A^\ast A}] < \infty$. The trace induces a norm on elements, $A$, in $\mathcal{T}(\mathcal{H})$ called the trace norm (or Schatten 1-norm) and is written as $\|A\|_1 := \operatorname{tr}[\sqrt{A^\ast A}]$. The second is the class of Hilbert-Schmidt operators, $\text{HS}(\mathcal{H})$, and satisfies that for all elements, $A$, we have $\operatorname{tr}[A^\ast A] < \infty$ and on this space the trace is an inner product -- which we will denote by $\langle \cdot | \cdot \rangle_\text{HS}$ -- turning the Hilbert-Schmidt class into a proper Hilbert space. Moreover, it turns out that $\mathcal{T}(\mathcal{H}) \subset \text{HS}(\mathcal{H}) \subset \mathcal{K}(\mathcal{H}) \subset \mathcal{B}(\mathcal{H})$ and it can be shown that via the canonical trace pairing $A \mapsto \operatorname{tr}[ \ \cdot \ A ]$, we have the following duality properties, $\mathcal{T}(\mathcal{H}) \cong \mathcal{K}(\mathcal{H})^\ast$ and $\mathcal{B}(\mathcal{H}) \cong \mathcal{T}(\mathcal{H})^\ast$. Lastly, in the case that $\mathcal{H} = L^2(\mathcal{X})$, we have that the space $\text{HS}(L^2(\mathcal{X}))$ is isomorphic to the integral kernels of $L^2(\mathcal{X} \times \mathcal{X})$.

The notion of a \textit{quantum state} can be lifted to classical mixtures using the set of density operators (sometimes just called \textit{states}) on $\mathcal{H}$, $\mathcal{D}(\mathcal{H})$, which is defined as the set of PSD operators, $\rho \geq 0$, with $\operatorname{tr}[\rho] = 1$, which automatically implies that $\rho$ is in $\mathcal{T}(\mathcal{H})$. By compactness and hermitianity, any element $\rho$ in $\mathcal{D}(\mathcal{H})$ allows a decomposition $\rho = \sum_{i = 0}^\infty \lambda_i |\psi_i\rangle \langle \psi_i|$, where the eigenvalues satisfy $\lambda_i \geq 0$ by PSD, $\sum_i \lambda_i = 1$ and $\{ \psi_i \}$ is an o.n.b. on $\mathcal{H}$. The purity of a state, $\rho$, is given by $\operatorname{tr}[\rho^2]$ and is bounded in $[(\dim{\mathcal{H}})^{-1},1]$. The set $\mathcal{D}(\mathcal{H})$ is moreover a convex, complete, separable metric space under the trace norm.

A \textit{(quantum) channel}, $\Phi$, from $\mathcal{T}(\mathcal{H}_A)$ to $\mathcal{T}(\mathcal{H}_B)$ is a superoperator satisfying the trace preserving property (TP), $\operatorname{tr}[\Phi[A]] = \operatorname{tr}[A]$ for any $A$ in $\mathcal{T}(\mathcal{H}_A)$, and the completely positive property (CP), which states that for any auxiliary finite dimensional Hilbert space, $\mathcal{R}$, and any $M \in \mathcal{T}(\mathcal{H}_A \otimes \mathcal{R})_+$ the operator $(\Phi \otimes I)[M]$ must be in $\mathcal{T}(\mathcal{H}_B \otimes \mathcal{R})_+$. The dual of a map $\Phi : \mathcal{T}(\mathcal{H}_A) \rightarrow \mathcal{T}(\mathcal{H}_B)$ is a map $\Phi^\ast: \mathcal{B}(\mathcal{H}_B) \rightarrow \mathcal{B}(\mathcal{H}_A)$ that is defined by $\operatorname{tr}[B \ \Phi[A]] = \operatorname{tr}[\Phi^\ast[B] \  A]$ for any $A$ in $\mathcal{T}(\mathcal{H}_A)$ and $B$ in $\mathcal{B}(\mathcal{H}_B)$. The adjoint of a CP map is CP -- mapping elements from $\mathcal{B}(\mathcal{H}_B \otimes \mathcal{R})_+$ to $\mathcal{B}(\mathcal{H}_A \otimes \mathcal{R})_+$ -- and the adjoint of a TP map has the property of being unital (U), $\Phi^\ast[I] = I$. Reversely, every \textit{normal} (meaning weak$^\ast$-continuous), UCP superoperator $\Phi^\ast : \mathcal{B}(\mathcal{H}_B) \rightarrow \mathcal{B}(\mathcal{H}_A)$ has a unique preadjoint $(\Phi^\ast)_\ast: \mathcal{T}(\mathcal{H}_A) \rightarrow \mathcal{T}(\mathcal{H}_B)$. In this sense, a CPTP map and its adjoint -- a normal UCP map -- form a pair analogous to the Schrödinger and Heisenberg pictures.

A channel can be described in a multitude of ways. The most important we will discuss, is the Stinespring representation. For every CP superoperator $\Phi^\ast: \mathcal{B}(\mathcal{H}_B) \rightarrow \mathcal{B}(\mathcal{H}_A)$, such a representation is given by a triple $(\pi, V, \mathcal{K})$, where $\pi : \mathcal{B}(\mathcal{H}_B) \rightarrow \mathcal{B}(\mathcal{K})$ is a unital *-homomorphism for some Hilbert space $\mathcal{K}$ and $V: \mathcal{H}_A \rightarrow \mathcal{K}$ is a bounded operator with $\| \Phi^\ast[I] \| = \| V \|^2$ such that $\Phi^\ast[B] = V^\ast \pi(B) V$ for all $B \in \mathcal{B}(\mathcal{H}_B)$. There is a certain sense in which we can have a smallest representation, called the minimal Stinespring representation. Such a representation $(\pi, V, \mathcal{K})$ satisfies that $\mathcal{K}$ is equal to the closed linear span of $\pi(\mathcal{B}(\mathcal{H}_B)) V \mathcal{H}_A$ and is unique in the sense any other minimal Stinespring representation is unitarily equivalent to it.

\section{Causal, unitary operators}
\label{sec:3}
It is a well-known result in operator and control theory that the time-invariant causal (TIC) operators on the frequency domain, can be identified with the Hardy on the upper half-plane $\mathbb{C}^+$ (see for example Theorem 10.3.5 in \cite{Staffans2005} for the Laplace domain version). The upper half-plane can be conformally mapped to the unit-disk using the Cayley transform, $c: \mathbb{C}^+ \rightarrow \mathbb{D}: z \mapsto (z - i)/(z + i)$. Hence, we only study operators on the unit-disk and remember that this can always be related to the frequency representation. In the following, let $\mathcal{U}, \mathcal{Y}$ be separable Hilbert spaces, let $\mathcal{X}$ be a Lebesgue measurable space and let $m$ be the Lebesgue measure.

\begin{definition}
    We denote $L^2(\mathbb{T}, \mathcal{U})$ as the set of strongly square integrable operator-valued functions on $\mathbb{T}$ with values in $\mathcal{U}$. We denote $L^\infty(\mathbb{T}, \mathcal{B}(\mathcal{U}, \mathcal{Y}))$ as the set of essentially bounded, WOT measurable operator-valued functions on $\mathbb{T}$ with values in $\mathcal{B}(\mathcal{U}, \mathcal{Y})$. Moreover $f \in H^2(\mathbb{T}, \mathcal{U})$ has a Fourier decomposition
    \[f_n := \int_\mathbb{T} \overline{x}^n f(x) \ dm(x)  \] 
    and $F \in H^\infty(\mathbb{T}, \mathcal{B}(\mathcal{U}, \mathcal{Y}))$ has a Fourier decomposition for all $u \in \mathcal{U}$
    \[ F_n u := \int_\mathbb{T} \overline{x}^n F(x) u  \ dm(x) \]
    with $n \in \mathbb{Z}$.
\end{definition}
    
\begin{definition}[Hardy space, first characterisation]
    Define $H^2(\mathbb{T}, \mathcal{U}) \subset L^2(\mathbb{T}, \mathcal{U})$ and $H^\infty(\mathbb{T}, \mathcal{B}(\mathcal{U}, \mathcal{Y})) \subset L^\infty(\mathbb{T}, \mathcal{B}(\mathcal{U}, \mathcal{Y}))$ as the subsets with no negative Fourier coefficients. 
\end{definition}

\begin{definition}[Hardy space, second characterisation]
    Define $H^2(\mathbb{D}, \mathcal{U})$ as the set of strongly analytic functions $\widetilde{f}: \mathbb{D} \rightarrow \mathcal{U}$ such that $ \sup_{r \in (0,1)} \| \widetilde{f}^r \|_{L^2(\mathbb{T}, \mathcal{U})} < \infty$ with $\widetilde{f}^r(x) = \widetilde{f}(r x)$. Define $H^\infty(\mathbb{D}, \mathcal{B}(\mathcal{U}, \mathcal{Y}))$ as the set of strongly analytic functions $\widetilde{F}: \mathbb{D} \rightarrow \mathcal{B}(\mathcal{U},\mathcal{Y})$ such that $\| \widetilde{F} \|_\infty < \infty$.
\end{definition}

There is a precise equivalence between the two types of Hardy classes, for which we need some machinery. 

\begin{definition}[Stolz region]
    The Stolz region $\Gamma_\alpha(x)$ with vertex $x \in \mathbb{T}$ and aperture $\alpha > 1$ is an open subset of $\mathbb{D}$ given by
    \[ \Gamma_\alpha(x) = \{ z \in \mathbb{D} \ | \ |z - x| < \alpha (1 - |z|) \} \text{.}\]
\end{definition}

\begin{definition}[Non-tangential boundary limit]
    Let $\tau \in \{ \text{strong}, \text{SOT}, \text{WOT} \}$, let $\widetilde{f} \in H^2(\mathbb{D}, \mathcal{U})$ and let $\widetilde{F} \in H^\infty(\mathbb{D}, \mathcal{B}(\mathcal{U}, \mathcal{Y}))$ . We say that $\widetilde{f}$ ($\widetilde{F}$) has $\tau$ non-tangential boundary limit $f(x)$ ($F(x)$) at $x \in \mathbb{T}$ if for every $\alpha > 1$ and $z \in \Gamma_\alpha(x)$ when $z \rightarrow x$ we have $\widetilde{f}(z) \rightarrow f(x)$ ($\widetilde{F}(z) \rightarrow F(x)$) in topology $\tau$.
\end{definition}

\begin{definition}[Poisson kernel]
    The Poisson kernel is given by 
    \[ P^r: \mathbb{T} \rightarrow \mathbb{R}_+: P^r(x) := (1-r^2)/|1-r x|^2 \]
    with $r \in [0,1)$. For $f \in L^2(\mathbb{T}, \mathcal{U})$, convolution by the Poisson kernel is given by
    \[ (f \ast P^r)(x) = \sum_{k \in \mathbb{Z}} f_k r^{|k|} x^k \text{.} \]
\end{definition}

The existence of non-tangential boundary limits in infinite dimensional Hilbert spaces is straightforward for $H^2$ spaces, but a bit more delicate in $H^\infty$. The following lemmas reflect this.

\begin{lemma}[Adapted from Theorem 3.11.7 in \cite{Nikolski2002}]
    \label{lem:Nikolski_simple}
    Let $\widetilde{f}$ in $H^2(\mathbb{D}, \mathcal{U})$. There exists the boundary function $f \in H^2(\mathbb{T}, \mathcal{U})$ defined a.e. by the existence of the strong non-tangential boundary limit $f(x) = \lim_{z \rightarrow x} \widetilde{f}(z)$ such that
    \[\widetilde{f}(r x) = (f \ast P^r)(x) \]
    for every $r \in [0,1)$ and $x \in \mathbb{T}$. Moreover $\widetilde{f} \mapsto f$ is a unitary between $H^2(\mathbb{D}, \mathcal{U})$ and $H^2(\mathbb{T}, \mathcal{U})$.
\end{lemma}

\begin{lemma}[Adapted from Theorem 3.11.10 in \cite{Nikolski2002}]
    \label{lem:Nikolski}
    Let $\widetilde{F}$ in $H^\infty(\mathbb{D}, \mathcal{B}(\mathcal{U}, \mathcal{Y}))$. There exists the boundary function $F \in H^\infty(\mathbb{T}, \mathcal{B}(\mathcal{U}, \mathcal{Y}))$ defined a.e. by the existence of the SOT non-tangential boundary limit $F(x) = \lim_{z \rightarrow x} \widetilde{F}(z)$ such that $F(\cdot) u$ is in $H^\infty(\mathbb{T}, \mathcal{Y})$ and
    \[\widetilde{F}(r x) u = (F u \ast P^r)(x) \]
    for every $u \in \mathcal{U}$ and $r \in [0,1)$, $x \in \mathbb{T}$. Moreover $\widetilde{F} \mapsto F$ is an isometric bijection between $H^\infty(\mathbb{D}, \mathcal{B}(\mathcal{U}, \mathcal{Y}))$ and $H^\infty(\mathbb{T}, \mathcal{B}(\mathcal{U}, \mathcal{Y}))$.
\end{lemma}

There are however some cases that we get stronger convergence and measurability.

\begin{lemma}[Adapted from Theorem 2.14 in \cite{Deutsch2010}]
    \label{lem:Deutsch}
    Let $T_n$ and $T$ be operators in $\mathcal{B}(\mathcal{U}, \mathcal{Y})$. Suppose $\mathcal{U}$ is finite dimensional, then $T_n$ converges to $T$ in the SOT if and only if it converges strongly.
\end{lemma}

\begin{lemma}[Adapted from p. 46 in \cite{Nikolski2002} ]
    \label{lem:Dunford}
    A function $F: \mathcal{X} \rightarrow \mathcal{B}(\mathcal{U}, \mathcal{Y})$ is strongly measurable if and only if it is WOT measurable and almost separable valued in $\mathcal{B}(\mathcal{U}, \mathcal{Y})$.
\end{lemma}

\begin{lemma}[Adapted from proposition 3.1.3 in \cite{Ende2020}]
    \label{lem:Ende}
    $\mathcal{B}(\mathcal{U}, \mathcal{Y})$ is separable if and only if $\mathcal{U}$ is finite dimensional or $\mathcal{Y}$ is finite dimensional.
\end{lemma}

\begin{lemma}
    \label{lem:Nikolski_adapted}
    Consider the situation in lemma \ref{lem:Nikolski} and suppose that additionally $\mathcal{U}$ (and not necessarily $\mathcal{Y}$) is finite dimensional, then $F$ is strongly measurable and $F(x) = \lim_{z \rightarrow x} \widetilde{F}(z)$ exists strongly.
    \begin{proof}
        Combine lemma \ref{lem:Deutsch}, \ref{lem:Dunford} and \ref{lem:Ende}.
    \end{proof}
\end{lemma}
The next definition tells us how the unitary aspect of our operators enters.
\begin{definition}[left inner, right inner, two-sided inner]
    Let $\widetilde{\Theta}$ be a function in $H^\infty(\mathbb{D}, \mathcal{B}(\mathcal{U}, \mathcal{Y}))$ and let $\Theta$ be its SOT non-tangential boundary function. Then we say that $\widetilde{\Theta}$ or $\Theta$ is left inner (right inner) if it is such that $\Theta^\ast(x) \Theta(x) = I$ ($\Theta(x) \Theta^\ast(x) = I$) a.e. $x$ in $\mathbb{T}$. $\widetilde{\Theta}$ or $\Theta$ is two-sided inner if it is both left inner and right inner.
\end{definition}
We need the following notions later on.

\begin{definition}[Flip]
    We define the flip as a map $\breve{\cdot}: L^\infty(\mathbb{T}, \mathcal{B}(\mathcal{U}, \mathcal{Y})) \rightarrow L^\infty(\mathbb{T}, \mathcal{B}(\mathcal{U}, \mathcal{Y})): F(x) \mapsto \breve{F}(x) := F(\overline{x})$.
\end{definition}

\begin{definition}[Reflection]
    We define the reflection as a map $\cdot^\sharp: L^\infty(\mathbb{T}, \mathcal{B}(\mathcal{U}, \mathcal{Y})) \rightarrow L^\infty(\mathbb{T}, \mathcal{B}(\mathcal{Y}, \mathcal{U})): F(x) \mapsto F^\sharp(x) = F^\ast(\overline{x})$. And thus $F^\sharp = \breve{F}^\ast$.
\end{definition}

\begin{remark}
    The reflection maps $H^\infty$ to $H^\infty$ and can thus be extended to the analytic functions.
\end{remark}

A special subset of Hardy class is the set of contraction-valued elements.

\begin{definition}[Schur-class functions]
    An operator-valued function, $\widetilde{S}: \mathbb{D} \rightarrow \mathcal{B}(\mathcal{U},\mathcal{Y})$, is called Schur-class if it is both strongly analytic on $\mathbb{D}$ and contraction-valued, $\| \widetilde{S}(z) \| \leq 1$ for all $z\in \mathbb{D}$. We denote the set of such Schur-class functions by $\mathcal{S}(\mathbb{D},\mathcal{B}(\mathcal{U}, \mathcal{Y}))$.
\end{definition}

\begin{remark}
    Schur-class functions, $\mathcal{S}(\mathbb{D},\mathcal{B}(\mathcal{U}, \mathcal{Y}))$, are elements of $H^\infty(\mathbb{D}, \mathcal{B}(\mathcal{U}, \mathcal{Y}))$.
\end{remark}

\begin{lemma}
    \label{lem:Schur_class}
    If $\widetilde{\Theta}$ in $H^\infty(\mathbb{D}, \mathcal{B}(\mathcal{U}, \mathcal{Y}))$ is left inner, then $\widetilde{\Theta}$ and $\widetilde{\Theta}^\sharp$ are Schur-class.
    \begin{proof}
        See appendix \ref{ap:proof_of_lemma14}.
    \end{proof}
\end{lemma}
\section{Causal quantum channels}
\label{sec:4}
\subsection{Schur multipliers}
\label{sec:4.1}
We recall the notion of a Schur multiplier and discuss some of its properties that will be relevant later.
\begin{definition}[Schur multiplier]
    Let $\varphi$ in $L^\infty(\mathcal{X} \times \mathcal{X})$. The map $M_\varphi: \text{HS}(L^2(\mathcal{X})) \rightarrow \text{HS}(L^2(\mathcal{X}))$ that takes integral kernels to integral kernels as $T(x,y) \mapsto (M_\varphi T)(x,y) := \varphi(x,y) T(x,y)$, is called a Schur multiplier with associated symbol $\varphi$.
\end{definition}

\begin{definition}[Transpose]
    We define the transpose $\cdot^\top: L^\infty(\mathcal{X} \times \mathcal{X}) \rightarrow L^\infty(\mathcal{X} \times \mathcal{X}): \varphi^\top(x,y) = \varphi(y,x)$.
\end{definition}
To consider Schur multipliers in the Heisenberg representation, we must allow action on a bigger space.
\begin{definition}[Measurable Schur multipliers, cf. lemma 3.1 in \cite{Duquet2022}]
    We will say that a Schur multiplier $M_\varphi: \text{HS}(L^2(\mathcal{X})) \rightarrow \text{HS}(L^2(\mathcal{X}))$, with $\varphi \in L^\infty(\mathcal{X} \times \mathcal{X})$ is measurable if it extends to a bounded, normal map $M_\varphi^\infty: \mathcal{B}(L^2(\mathcal{X})) \rightarrow \mathcal{B}(L^2(\mathcal{X}))$.
\end{definition}

\begin{remark}
    Any Schur multiplier that is bounded as a map $M_\varphi : \mathcal{T}(L^2(\mathcal{X})) \rightarrow \mathcal{T}(L^2(\mathcal{X}))$ naturally corresponds to a measurable Schur multiplier with symbol $\varphi^\top$.
\end{remark}

From now on, we will omit the $\cdot^\infty$ for the measurable Schur multipliers when it is clear that we mean the extension.

\begin{lemma}[Adapted from Theorem 3.2 in \cite{Duquet2022}]
    \label{lem:meas_schur_mult}
    $M_\varphi$ with $\varphi \in L^\infty(\mathcal{X} \times \mathcal{X})$ is a measurable Schur multiplier if and only if there exists a Hilbert space $\mathcal{H}$ and two functions $\alpha$ and $\beta$ in $L^\infty(\mathcal{X}, \mathcal{H})$ such that
    \[\varphi(x,y) = \langle \alpha(x) | \beta(y) \rangle_{\mathcal{H}} \text{ a.e. on } \mathcal{X} \times \mathcal{X} \text{.} \]
\end{lemma}

\begin{lemma}[Adapted from Theorem 3.5 in \cite{Duquet2022}]
    \label{lem:unital_of_schur}
    Let $M_\varphi$ be a measurable Schur multiplier with symbol $\varphi$ in $L^\infty(\mathcal{X} \times \mathcal{X})$. Then $M_\varphi$ is unital if and only if for $\alpha, \beta$ satisfying lemma \ref{lem:meas_schur_mult}
    \[
        \langle \alpha(x) | \beta(x) \rangle_{\mathcal{H}} = 1 \quad \text{a.e. on } \mathcal{X} \text{.}
    \]
\end{lemma}

\begin{definition}[Positive Schur multiplier]
    A Schur multiplier $M_\varphi$ with symbol $\varphi \in L^\infty(\mathcal{X} \times \mathcal{X})$ is said to be positive whenever
    \[ M_\varphi( \text{HS}(L^2(\mathcal{X} ) )_+ ) \subseteq \text{HS}(L^2(\mathcal{X} ))_+ \text{.}\]
    
\end{definition}

\begin{lemma}[Adapted from corollary 4.5 in \cite{Steen2012}]
    \label{lem:positivity_and_factorization}
    Let $M_\varphi$ be a measurable Schur multiplier with symbol $\varphi$ in $L^\infty(\mathcal{X} \times \mathcal{X})$. Then the following are equivalent:
    \begin{enumerate}
    \small
        \item $M_\varphi$ is positive.
        \item $M_\varphi$ is completely positive.
        \item There exists an essentially bounded, measurable map $a: \mathcal{X} \to \ell^2$ such that
              \[
                  \varphi(x,y) = \langle a(x) | a(y) \rangle_{\ell^2}, \quad \text{a.e. on } \mathcal{X} \times \mathcal{X} \text{.}
              \]
    \end{enumerate}

    \begin{remark}
        If $M_\varphi$ is positive, there exists an essentially bounded, measurable function $a:\mathcal{X} \to\ell^2$ such that
        \[
            \varphi(x,y) = \langle a(x) | a(y) \rangle_{\ell^2} \quad \text{a.e.,}
        \]
        which provides a Stinespring-type representation for $M_\varphi$, proving complete positivity.
    \end{remark}
\end{lemma}

\begin{lemma}[Adapted from proposition 4.6 in \cite{Steen2012}]
    \label{lem:minimal_stinespring_rep}
    Let $M_\varphi$ with $\varphi$ in $L^\infty(\mathcal{X} \times \mathcal{X})$ be a positive, measurable Schur multiplier and $a: \mathcal{X} \rightarrow \mathcal{Y}$ with $\mathcal{Y} = \ell^2$ or $\mathcal{Y} = \mathbb{C}^d$ such that (3) in lemma \ref{lem:positivity_and_factorization} holds. Then the following are equivalent
    \begin{enumerate}
        \small
        \item $M_\varphi T = \sum_{i=1}^{\dim \mathcal{Y}} M_{a_i}^\ast T M_{a_i}, T \in \text{HS}(L^2(\mathcal{X}))$ is a minimal Stinespring representation of $M_\varphi$.
        \item for each null set $M \subseteq \mathcal{X}$, the set $\{a(x) \ | \ x \in \mathcal{X} \setminus M \}$ has dense linear span in $\mathcal{Y}$.
    \end{enumerate}

    \begin{remark}
        Note that in \cite{Steen2012} the proof of proposition 4.6 depends on lemma 3.1 and that finite dimensionality is a special case herein.
    \end{remark}

\end{lemma}
\subsection{Schur-class kernels}
\label{sec:4.2}
In the following section we define the notion of a Schur-class kernel. We show that these objects naturally arise as classical mixtures of Schur-class functions. Then we prove that Schur-class kernels viewed as multiplicative maps are completely positive measurable Schur multipliers by providing a factorization of the kernel using reproducing kernel Hilbert spaces (RKHS). Next, we show that this factorization corresponds to a Stinespring representation that is in fact minimal. If unitality of the Schur-class kernel is moreover demanded, we show that the vectors in the factorization correspond with left inner functions.

\begin{definition}[Positive semidefinite kernel]
    A Hermitian function, $K: \mathcal{X} \times \mathcal{X} \rightarrow \mathbb{C}$, is called a positive semidefinite kernel, if for each set $\{a_i\}_{i=1}^n \subset \mathbb{C}$ and $\{x_i\}_{i=1}^n \subseteq \mathcal{X}$, where $n \in \mathbb{N}$, we have
    \[ \sum_{i,j=1}^n \overline{a_i} a_j \ K(x_i, x_j) \geq 0 \text{.}\]
\end{definition}

\begin{definition}[Sesquianalytic]
    A function $f: A \times B \rightarrow \mathbb{C}$ where $A, B$ are open subsets of $\mathbb{C}$, is called sesquianalytic if it is anti-analytic as a map $f(\cdot, b): a \mapsto f(a,b)$ for all $b$ in $B$ and analytic as a map $f(a, \cdot): b \mapsto f(a,b)$ for all $a$ in $A$.
\end{definition}

\begin{definition}[Schur-class kernels]
    Let $\varphi \in L^\infty(\mathbb{T} \times \mathbb{T})$. If there exists a function $\widetilde{\varphi}: \mathbb{D} \times \mathbb{D} \rightarrow \mathbb{C}$ such that
    \begin{enumerate}
    \small
        \item $\widetilde{\varphi}$ is a positive semidefinite kernel.
        \item $\widetilde{\varphi}$ is sesquianalytic.
        \item $\widetilde{\varphi}$ is bounded in supremum norm.
        \item there exists some conull set $E \subseteq \mathbb{T} \times \mathbb{T}$ such that if $z \rightarrow x$ and $w \rightarrow y$ independently and non-tangentially for $(x,y) \in E$, then $\widetilde{\varphi}(z,w) \rightarrow \varphi(x,y)$.
    \end{enumerate}
    then $\varphi$ is called a Schur-class kernel and $\widetilde{\varphi}$ is called its extension.
\end{definition}

\begin{remark}
    In the definition of Schur-class kernel, requirement 2 and 3 may be replaced by the demand that $(z,w) \mapsto \widetilde{\varphi}(\overline{z}, w)$ is an $H^\infty(\mathbb{D} \times \mathbb{D})$ function.
\end{remark}

The next Theorem shows that Schur-class kernels naturally arise as classical mixtures of Schur-class functions. 

\begin{theorem}[Schur-class kernels from Schur-class functions]
    \label{thm:1}
    Let $(R, \mathcal{R}, \mu)$ be a finite measure space. Let $F: R \rightarrow \mathcal{S}(\mathbb{D},\mathcal{B}(\mathcal{U}, \mathcal{Y})): \scriptr \mapsto \widetilde{S}_\scriptr$ be a function with values in the Schur-class functions, such that $\scriptr \mapsto \widetilde{S}_\scriptr(z)$ is WOT measurable for all $z \in \mathbb{D}$. Then for any orthogonal projection $P$ on a finite dimensional subspace of $\mathcal{U}$ we have that
    \[ \widetilde{\varphi}(z,w) := \int_R \operatorname{tr}[ \widetilde{S}_\scriptr(w) \ P \ \widetilde{S}_\scriptr^\ast(z) ] \ d\mu(\scriptr) \]
    is the extension of a Schur-class kernel.

    \begin{proof}
        For the full proof see appendix \ref{ap:proof_of_thm1}. Sketch: $\widetilde{\varphi}$ inherits its properties from $\widetilde{S}_\scriptr$ and the finite dimensionality of $P$ ensures the justification of the trace. The points at which $\widetilde{\varphi}$ does not have a boundary limit can be shown to have measure zero, using Fubini-Tonelli.
    \end{proof}

\end{theorem}

\begin{definition}[Operator representation of a vector]
    Define $[\cdot]: \mathcal{Y} \rightarrow \mathcal{B}(\mathbb{C}, \mathcal{Y}): y \mapsto [y]$ as the map that takes vectors to operators.
\end{definition}

\begin{theorem}[Factorization of Schur-class kernels]
    \label{thm:Fact_of_kernels}
    Let $\varphi \in L^\infty(\mathbb{T} \times \mathbb{T})$ be a Schur-class kernel. Then there exists a separable Hilbert space, $\mathcal{H}_{\widetilde{\varphi}}$, and a bounded, strongly analytic vector-valued function $\widetilde{\psi}: \mathbb{D} \rightarrow \mathcal{H}_{\widetilde{\varphi}}$ such that $\widetilde{\varphi}(z,w) = \langle {\widetilde{\psi}}(z) | {\widetilde{\psi}}(w) \rangle_{\mathcal{H}_{\widetilde{\varphi}}}$ pointwise on $\mathbb{D} \times \mathbb{D}$. Moreover, ${\widetilde{\psi}}$ has an essentially bounded, strongly measurable boundary function $\psi$ such that $\varphi(x,y) = \langle \psi(x) | \psi(y) \rangle_{\mathcal{H}_{\widetilde{\varphi}}}$ a.e. on $\mathbb{T} \times \mathbb{T}$.
    \begin{proof}
        By the Moore-Aronszajn Theorem, there exists a reproducing kernel Hilbert space, $\mathcal{H}_{\widetilde{\varphi}}$ of anti-analytic functions with kernel $\widetilde{\varphi}$, formed by the closed linear span of functions $\widetilde{\varphi}(\cdot, z): w \mapsto \widetilde{\varphi}(w, z)$ with domain $\mathbb{D}$. Define the vector-valued function $\widetilde{\psi}: \mathbb{D} \rightarrow \mathcal{H}_{\widetilde{\varphi}} : z \mapsto \widetilde{\varphi}(\cdot, z)$, which by construction satisfies $\langle \widetilde{\psi}(z) | \widetilde{\psi}(w) \rangle_{\mathcal{H}_{\widetilde{\varphi}}} = \widetilde{\varphi}(z, w)$. Boundedness of ${\widetilde{\psi}}$ follows from
        \[ \| {\widetilde{\psi}} \|_\infty = \sup_{z \in \mathbb{D}} \|{\widetilde{\psi}}(z) \| = \sup_{z \in \mathbb{D}} \sqrt{\langle \widetilde{\varphi}(\cdot, z)  | \widetilde{\varphi}(\cdot, z)  \rangle_{\mathcal{H}_{\widetilde{\varphi}}}} = \sup_{z \in \mathbb{D}} \sqrt{\widetilde{\varphi}(z, z)} \leq \sqrt{\| \widetilde{\varphi} \|_\infty}  < \infty\]
        because $\widetilde{\varphi}$ is bounded. Now, $\widetilde{\varphi}$ being sesquianalytic\footnote{By Hartog's Theorem separate analyticity and joint analyticity coincide.} implies that $\widetilde{\varphi}$ is jointly continuous. Together with the fact that $\mathbb{D}$ is separable, we find (lemma 4.33 in \cite{Steinwart2008}) that $\mathcal{H}_{\widetilde{\varphi}}$ is separable. Then, choose $f \in \mathcal{H}_{\widetilde{\varphi}}$ arbitrary and let $\{ f^{(n)} \} \subset \operatorname{span} \{ \widetilde{\varphi}(\cdot, z) \ | \ z \in \mathbb{D}\}$ be a limiting sequence of $f$ which for each $f^{(n)}$ can be written as
        \[ f^{(n)}(w) = \sum_{i = 1}^{N_n} a_i^{(n)} \widetilde{\varphi}(w, z_i)\]
        with $N_n < \infty$ and sets $\{z_i \} \subset \mathbb{D}$ and $\{a^{(n)}_i \}  \subset \mathbb{C}$. We find that
        \[ \sup_{z \in \mathbb{D}} | \langle f^{(n)} - f | {\widetilde{\psi}}(z)  \rangle_{\mathcal{H}_{\widetilde{\varphi}}} | \leq  \| f^{(n)} - f \| \| {\widetilde{\psi}} \|_\infty \rightarrow 0 \]
        implies that $\langle f^{(n)} | {\widetilde{\psi}}(z)  \rangle_{\mathcal{H}_{\widetilde{\varphi}}}$ uniformly converges to $\langle f | {\widetilde{\psi}}(z) \rangle_{\mathcal{H}_{\widetilde{\varphi}}}$ and each
        \[ \langle f^{(n)} | {\widetilde{\psi}}(z) \rangle_{\mathcal{H}_{\widetilde{\varphi}}} = \sum_{i=1}^{N_n} \overline{a^{(n)}_i} \widetilde{\varphi}(z_i, z) \]
        is analytic. By Weierstrass' Theorem on uniform convergence this means that ${\widetilde{\psi}}$ is weakly analytic on $\mathbb{D}$. Because $\widetilde{\psi}$ is (globally) bounded, Theorem 6.4 in \cite{Reed1980} implies that ${\widetilde{\psi}}(z)$ is strongly analytic on $\mathbb{D}$. Next, by lemma \ref{lem:Nikolski_adapted} the boundary limit of the column $[\widetilde{\psi}]$ in $H^\infty(\mathbb{D}, \mathcal{B}(\mathbb{C}, \mathcal{H}_{\widetilde{\varphi}}))$ exists for strong non-tangential approach on a conull set $G \subseteq \mathbb{T}$. The resulting boundary function, $[\psi]$, is strongly measurable and essentially bounded on $\mathbb{T}$. Lastly, let $E \subseteq \mathbb{T} \times \mathbb{T}$ be conull set such that $\varphi$ is the boundary function of $\widetilde{\varphi}$ on $E$. Consider that
        \[ \begin{split}
                |\varphi(x_1, x_2) - \langle \psi(x_1) | \psi(x_2) \rangle_{\mathcal{H}_{\widetilde{\varphi}}} | = | & \varphi(x_1, x_2) - \widetilde{\varphi}(z_1, z_2) + \widetilde{\varphi}(z_1, z_2) \\ - \langle &\psi(x_1) | \psi(x_2) \rangle_{\mathcal{H}_{\widetilde{\varphi}}}  + \langle \widetilde{\psi}(z_1)| \widetilde{\psi}(z_2) \rangle_{\mathcal{H}_{\widetilde{\varphi}}} - \langle \widetilde{\psi}(z_1)| \widetilde{\psi}(z_2) \rangle_{\mathcal{H}_{\widetilde{\varphi}}} | \\
                \leq |                                                                                           & \varphi(x_1, x_2) - \widetilde{\varphi}(z_1 , z_2) |                          \\ + | &\langle \psi(x_1) | \psi(x_2) \rangle_{\mathcal{H}_{\widetilde{\varphi}}}  - \langle \widetilde{\psi}(z_1) | \widetilde{\psi}(z_2) \rangle_{\mathcal{H}_{\widetilde{\varphi}}} | \\
                \leq |                                                                                           & \varphi(x_1, x_2) - \widetilde{\varphi}(z_1 , z_2) |                          \\ + \| & \psi(x_1) - \widetilde{\psi}(z_1) \| \| \psi \|_\infty + \| \widetilde{\psi} \|_\infty \| \psi(x_2) - \widetilde{\psi}(z_2) \|
            \end{split}\]
        which goes to zero as $z_1 \rightarrow x_1$ and $z_2 \rightarrow x_2$ non-tangentially for $(x_1,x_2) \in E \cap (G \times G)$.
    \end{proof}
\end{theorem}

\begin{remark}[Upper bound on the dimension of the RKHS]
    \label{rem:Papadakis}
    Suppose that for a Schur-class kernel, $\varphi$, and some $a_i: \mathbb{D} \rightarrow \mathbb{C}$ for $i=1, \dots, n$ the series
    \[ \widetilde{\varphi}(z,w) = \sum_{i=1}^n \overline{a_i(z)} a_i(w) \]
    converges pointwise. Then by Papadakis' Theorem (see Theorem 2.10 and exercise 3.7 in \cite{Paulsen2016}) $\{ \overline{a_i(z)} \}_{i=1}^n\subseteq \mathcal{H}_{\widetilde{\varphi}}$ is a Parseval frame of $\mathcal{H}_{\widetilde{\varphi}}$ and thus $\dim \mathcal{H}_{\widetilde{\varphi}} \leq n$.
\end{remark}

\begin{corollary}[Minimal Stinespring representation]
    \label{cor:Stinespring_rep}
    Consider the situation as in Theorem \ref{thm:Fact_of_kernels}. Then $M_\varphi$ is a measurable, CP Schur multiplier and there exists a Hilbert space, $\mathcal{K}_d = \mathbb{C}^d$ if $d<\infty$ and $\mathcal{K}_d = \ell^2$ if $d = \infty$, and an operator-valued function $[\widetilde{v}]$ in $H^\infty(\mathbb{D}, \mathcal{B}(\mathbb{C}, \mathcal{K}_d))$ such that for its boundary function $[v]$ we have for $T$ in $\text{HS}(L^2(\mathbb{T}))$ and a.e. $x,y \in \mathbb{T}$ that
    \[ (M_\varphi T)(x,y) \cong (I \otimes [v](x) )^\ast \ (T(x,y) \otimes I) \ (I \otimes [v](y) ) \]
    is a minimal Stinespring representation of $M_\varphi$.
    \begin{proof}
        $\mathcal{H}_{\widetilde{\varphi}}$ permits a countable, orthonormal basis $\{e_n\}_{n = 1}^d$ with $d := \dim \mathcal{H}_{\widetilde{\varphi}} \in [0, \infty]$. Then, define the unitary $U: \mathcal{H}_{\widetilde{\varphi}} \rightarrow \mathcal{K}_d: f \mapsto \{ \langle e_n | f\rangle_{\mathcal{H}_{\widetilde{\varphi}}} \}_{n = 1}^d$ and observe that under this map $\widetilde{v} := U(\widetilde{\psi})$ is a bounded, strongly analytic, vector-valued function with essentially bounded, strongly a.e. measurable boundary function $v$ that agrees with $U(\psi)$ by unitarity. Lemma \ref{lem:meas_schur_mult} then implies that $M_\varphi$ is a measurable Schur multiplier and lemma \ref{lem:positivity_and_factorization} implies that $M_\varphi$ is CP.

        Because $\mathcal{H}_{\widetilde{\varphi}} = \operatorname{clos \ span}\{ \widetilde{\varphi}(\cdot, z) \ | \ z \in \mathbb{D}\}$, we have that $\mathcal{K}_d \equiv \operatorname{clos \ span}\{ \widetilde{v}(z) \ | \ z \in \mathbb{D}\}$ by unitarity. Take now any $a$ in $\mathcal{K}_d$ and suppose that $\langle a | v(x) \rangle = 0$ on some conull set $\mathcal{X} \subseteq \mathbb{T}$. Now, $\langle a | v(x) \rangle$ on $\mathcal{X}$ extends to the analytic function $\langle a | \widetilde{v}(z) \rangle$ which means that $\langle a | \widetilde{v}(z) \rangle = 0$ for all $z$ in $\mathbb{D}$ by boundary uniqueness. This means that $a \equiv 0$ and thus it must hold that $\mathcal{K}_d \equiv \operatorname{clos \ span}\{ v(x) \ | \ x \in \mathcal{X} \}$. By lemma \ref{lem:minimal_stinespring_rep} this implies that $v$ induces a minimal Stinespring representation.
    \end{proof}
\end{corollary}

It is a known fact that Stinespring representations that are minimal are essentially unique (see also proposition 4.2 in \cite{Paulsen2016}). The following corollary gives insight in how a TIC quantum channel increases rank.

\begin{corollary}[Spectral decomposition]
    \label{cor:spec_decomp}
    Let $\varphi$ be a Schur-class kernel and let $\psi \in L^2(\mathbb{T})$ be any non-zero vector. Then there exist $\{ h_i \}_{i=1}^d \subset H^\infty(\mathbb{T})$ and values $\{ \lambda_i \}_{i=1}^d \subset \mathbb{R}_+$ with $d = \dim \mathcal{H}_{\widetilde{\varphi}}$ such that $\{ u_i := M_{h_i} \psi\}_{i=1}^d \subset L^2(\mathbb{T})$ is an orthonormal set of functions and a.e.
    \[ \varphi(x,y) \overline{\psi(x)} \psi(y) = \sum_{i=1}^d \lambda_i \overline{u_i(x)} u_i(y) \text{.} \]
    \begin{proof}
        See appendix \ref{ap:proof_of_spec_decomp}.
    \end{proof}

\end{corollary}

The additional requirement of unitality turns the earlier defined Schur-class kernel into the desired object, the TIC quantum channel. The following corollary further classifies the vector, $v$.

\begin{corollary}[Unital Schur-class kernels]
    \label{cor:unital_real_schur_mult}
    Consider the situation as in Theorem \ref{thm:Fact_of_kernels} and suppose additionally that $M_\varphi$ is unital, then $[\widetilde{v}]$ is a left inner operator-valued function.
    \begin{proof}
        By corollary \ref{cor:Stinespring_rep}, we know that $M_\varphi$ is a measurable Schur multiplier and by lemma \ref{lem:unital_of_schur} we know that
        \[ \| [v](x) \|^2 = \langle v(x) | v(x) \rangle = 1 \]
        a.e. $x \in \mathbb{T}$, which implies that $[\widetilde{v}]$ is a left inner operator-valued function.
    \end{proof}
\end{corollary}

\subsection{Complementary factors}
\label{sec:4.3}
In this section we discuss the existence of a completion of left inner functions to two-sided inner functions in potentially infinite dimensions. We show that using a realization of the left inner function, the demand for completability can be related to strong stability. Moreover, the Julia embedding of such a realization is a natural candidate for the smallest of such completions, which turns out to be the case. We then further specify to the case of finite dimensional, rational inner functions and show that we can in fact compute the completion with the developed tools. For convenience, we will omit the $\widetilde{\cdot}$ for the distinction between functions on the boundary and the interior.

\begin{definition}[Toeplitz and Hankel operators]
    Let $F \in L^\infty(\mathbb{T}, \mathcal{B}(\mathcal{U}, \mathcal{Y}))$, let $P_+^\mathcal{U}, P_-^\mathcal{U}$ be orthoprojections on $H^2(\mathbb{T}, \mathcal{U})$ and $L^2(\mathbb{T}, \mathcal{U}) \ominus H^2(\mathbb{T}, \mathcal{U})$ respectively and let $J^\mathcal{U}: L^2(\mathbb{T}, \mathcal{U}) \rightarrow L^2(\mathbb{T}, \mathcal{U}): f(z) \mapsto \overline{z} f(\overline{z})$ be the unitary involution mapping $L^2(\mathbb{T}, \mathcal{U}) \ominus H^2(\mathbb{T}, \mathcal{U})$ to $H^2(\mathbb{T}, \mathcal{U})$. Then the Toeplitz operator is defined as 
    \[ T_F = P_+^\mathcal{Y} M_F P_+^\mathcal{U}  \]
    and the Hankel operator as 
    \[ H_F = J^\mathcal{Y} P_-^\mathcal{Y} M_F P_+^\mathcal{U} \text{.} \]
\end{definition}
The following Toeplitz-Hankel identities are needed, which are easily verified.
\begin{lemma}
    \label{lem:th_identities}
    Let $F \in L^\infty(\mathbb{T}, \mathcal{B}(\mathcal{V}, \mathcal{Y}))$ then
    \begin{enumerate}
        \item $T_F^\ast = T_{F^\ast}$, $H_F^\ast = H_{F^\sharp} $.
        \item $H_F T_G = H_{F G}$ if $G \in H^\infty(\mathbb{T}, \mathcal{B}(\mathcal{U}, \mathcal{V}))$.
        \item $H_{G F} = T_{G^\sharp}^\ast H_F$ if $G \in H^\infty(\mathbb{T}, \mathcal{B}(\mathcal{Y}, \mathcal{Z}))$.
        \item $T_F T_G = T_{FG} - H_{F^\ast}^\ast H_G$ if $G \in L^\infty(\mathbb{T}, \mathcal{B}(\mathcal{U}, \mathcal{V}))$.
    \end{enumerate}
\end{lemma}

\begin{definition}[Shift operator, backward shift operator]
    The shift operator is a map
    \[ \mathbf{S}_\mathcal{U} : H^2(\mathbb{D}, \mathcal{U}) \rightarrow H^2(\mathbb{D}, \mathcal{U}): f(z) \mapsto z f (z) \]
    and its adjoint is called the backward shift operator
    \[ \mathbf{S}_\mathcal{U}^\ast : f(z) \mapsto \frac{f(z) - f(0)}{z} \text{.}\]
\end{definition}

\begin{lemma}[Adapted from \cite{Curto2021}]
    \label{lem:commutator_of_shift}
    Let $F \in L^\infty(\mathbb{T}, \mathcal{B}(\mathcal{U}, \mathcal{Y}))$ then
    \[ H_F \mathbf{S}_\mathcal{U} = \mathbf{S}_\mathcal{Y}^\ast H_F  \text{.}\]
    \begin{proof}
        Combine 2. and 3. in lemma \ref{lem:th_identities}.
    \end{proof}
\end{lemma}

\begin{lemma}[Adapted from lemma 2.7 in \cite{Curto2021}]
    \label{lem:curto_4.5}
    Let $\Theta \in H^\infty(\mathbb{T}, \mathcal{B}(\mathcal{U},\mathcal{Y}))$ be an operator-valued left inner function. By the Beurling-Lax-Halmos (BLH) Theorem, we may write
    \[ \ker \left( M_{\Theta^\ast}|_{H^2(\mathbb{T}, \mathcal{Y})} \right) = \Omega H^2(\mathbb{T}, \mathcal{V}) \]
    for some left inner function $\Omega: \mathbb{D} \rightarrow \mathcal{B}(\mathcal{V},\mathcal{Y})$. Define the BLH complementary factor of $\Theta$ by
    \[ \Theta^c_{\text{BLH}} := \text{left-g.c.d} \{ [g]^i : g \in \ker \Theta^\ast \}\]
    with $[g]^i$ the left inner part of $[g]$. Then we have
    \begin{enumerate}
    \small
        \item $\Omega = \Theta^c_{\text{BLH}}$.
        \item $\begin{pmatrix} \Theta &  \Theta^c_{\text{BLH}} \end{pmatrix}$ is a left inner function with values in $\mathcal{B}(\mathcal{U} \oplus \mathcal{V}, \mathcal{Y})$.
        \item $\ker H_{\Theta^\ast} = \begin{pmatrix} \Theta & \Theta^c_{\text{BLH}} \end{pmatrix} H^2(\mathbb{T}, \mathcal{U} \oplus \mathcal{V})$.
    \end{enumerate}
    where $\begin{pmatrix} \Theta & \Theta^c_{\text{BLH}}\end{pmatrix}$ is a $1 \times 2$ operator matrix.
\end{lemma}
These BLH complementary factors always exist and are moreover unique up to constant unitary right factor.
\begin{lemma}[Adapted from corollary 2.25 in \cite{Curto2021}] \label{lem:curto_4.20}    
    Let $\Theta \in H^\infty(\mathbb{T}, \mathcal{B}(\mathcal{U},\mathcal{Y}))$ be left inner. Then
    \[ \breve{\Theta} \text{ is of bounded type} \iff \begin{pmatrix} \Theta & \Theta^c_{\text{BLH}} \end{pmatrix} \text{ is two-sided inner} \text{.}\]
\end{lemma}
That $\breve{\Theta}$ be \textit{of bounded type} or \textit{Nevanlinna-class} for finite dimensional $\mathcal{U}, \mathcal{Y}$ requires that each entry of the matrix $\breve{\Theta}$ coincides with the quotient of bounded, analytic functions. We moreover have the restrictive result.
\begin{lemma}[Adapted from corollary 2.26 in \cite{Curto2021}] \label{lem:curto_4.21}  
    Let $\Theta$ be an inner function with values in $\mathcal{B}(\mathcal{U}, \mathcal{Y})$. Then $\begin{pmatrix} \Theta & \Theta^c \end{pmatrix}$ is two-sided inner for some inner function $\Theta^c$ with values in $\mathcal{B}(\mathcal{V}, \mathcal{Y})$ if and only if $\breve{\Theta}$ is of bounded type.
\end{lemma}
We will refer to left inner functions that have such a complementary factor $\Theta^c$ as \textit{completable}. The previous lemma also has the consequence that if $\Theta^c$ is a complementary factor, then $\Theta^c_\text{BLH}$ is a left inner divisor of $\Theta^c$. Hence we will call $\Theta^c_\text{BLH}$ and any complementary factor related by a constant right unitary factor the \textit{minimal} complementary factor. Now we transfer these results to the realization spaces.

\begin{definition}[Realization and characteristic function, adapted from \cite{Lilleberg2019}]
    Let $\Sigma: \mathcal{H} \oplus \mathcal{U} \rightarrow \mathcal{H} \oplus \mathcal{Y}$ be an operator matrix given by
    \[ \Sigma = \begin{pmatrix}
            A & B \\
            C & D\end{pmatrix} \]
    where $A,B,C,D$ are respectively called the main, control, observation and feedthrough operators and $\mathcal{H}$ is a separable Hilbert space, called the state-space. Moreover, let $F \in H^\infty(\mathbb{D},\mathcal{B}(\mathcal{U}, \mathcal{Y}))$ be some operator valued function. Define the characteristic function of $\Sigma$ as
    \[ \theta_\Sigma(z) := D + z C (I - z A)^{-1} B \]
    for $z \in \mathbb{D}$. We say that $\Sigma$ is a realization of $F$ if $\theta_\Sigma = F$.
\end{definition}

\begin{definition}[Controllable, observable, simple, minimal, passive and conservative realization, adapted from \cite{Lilleberg2019}]
    A realization $\Sigma: \mathcal{H} \oplus \mathcal{U} \rightarrow \mathcal{H} \oplus \mathcal{Y}$ is called controllable if
    \[ \mathcal{H} = \operatorname{clos \ span} \{ \operatorname{ran} A^n B \ | \ n \in \mathbb{N}\} \]
    observable if
    \[ \mathcal{H} = \operatorname{clos \ span} \{ \operatorname{ran} (A^\ast)^n C^\ast \ | \ n \in \mathbb{N}\}\]
    and simple if
    \[ \mathcal{H} = \operatorname{clos \ span} \{ \operatorname{ran} A^n B, \operatorname{ran} (A^\ast)^m C^\ast \ | \ n,m \in \mathbb{N}\} \text{.}\]
    The realization is called minimal if it is both controllable and observable. A realization is called passive, if it is contractive and it is called conservative if it is unitary.
\end{definition}
\begin{remark}
    \label{rem:simple}
    Any realization that is observable or controllable, is automatically simple.
\end{remark}

\begin{definition}[de Branges-Rovnyak/ model space, canonical realization, adapted from \cite{Ball2006}]
    \label{def:can_real}
    Let $S$ in $\mathcal{S}(\mathbb{D},\mathcal{B}(\mathcal{U}, \mathcal{Y}))$ be a Schur-class function and let $K_S$ be the kernel
    \[ K_S(z,w) := \frac{1 - S(z) S^\ast(w)}{1 - z \bar{w}} \text{.}\]
    Define $\mathcal{H}(S)$ as the reproducing kernel Hilbert space generated by $K_S$, which is the so-called de Branges-Rovnyak or model space tied to $S$. We define the canonical realization of $S$ to be the operator that maps $\mathcal{H}(S) \oplus \mathcal{U} \rightarrow \mathcal{H}(S) \oplus \mathcal{Y}$ given by the operator matrix
    \[U_S = \begin{pmatrix}
            A_S & B_S \\
            C_S & D_S
        \end{pmatrix} := \begin{pmatrix}
            R_0  & \tau \\
            e(0) & S(0)
        \end{pmatrix} \]
    where $R_0 =  \mathbf{S}_\mathcal{Y}^\ast|_{\mathcal{H}(S)}$ is the backward-shift operator, $\tau(z) = (S(z) - S(0))/z$ and $e(0)$ is the evaluation-at-0 map.
\end{definition}

\begin{lemma}[Adapted from Theorem 1.3 in \cite{Ball2006}]
    \label{lem:Ball}
    Let $S$ and $U_S$ be as in definition \ref{def:can_real}. Then
    \begin{enumerate}
    \small
        \item $U_S$ realizes $S$.
        \item $U_S$ is coisometric.
        \item $U_S$ is observable.
        \item Any other coisometric, observable realization of $S$ is unitarily equivalent to $U_S$.
    \end{enumerate}
\end{lemma}

There is a different way that the de Branges-Rovnyak space can be generated which greatly simplifies in the case of inner functions.

\begin{lemma}[Adapted from proposition 5 in \cite{Ball2014}]
    \label{lem:prop_36}
    Let $S$ in $\mathcal{S}(\mathbb{D},\mathcal{B}(\mathcal{U}, \mathcal{Y}))$ be a Schur-class function, let $\mathcal{H}(S)$ be its model space and let $T_S$ be its Toeplitz operator. Then
    \[ \mathcal{H}(S) = \operatorname{ran}(I - T_S T_S^\ast)^{1/2} \]
    together with lifted norm
    \[ \| (I - T_S T_S^\ast)^{1/2} f \|_{\mathcal{H}(S)} = \| Q f \|_{H^2(\mathbb{D}, \mathcal{Y})}, \quad f \in H^2(\mathbb{D}, \mathcal{Y}) \]
    where $Q$ is the orthoprojector onto $(\operatorname{ker}(I - T_S T_S^\ast))^\perp$.
\end{lemma}

\begin{lemma}[Model spaces of right inner functions]
    \label{lem:mod_space}
    Let $\Theta$ be a right inner function in $H^\infty(\mathbb{D}, \mathcal{B}(\mathcal{U}, \mathcal{Y}))$. Then 
    \[ \mathcal{H}(\Theta) = \operatorname{ran} (H_{\breve{\Theta}}) \text{.} \]
    \begin{proof}
    Invoking 1. and 4. in lemma \ref{lem:th_identities} and right innerness we have
    \[ H_{\Theta^\ast}^\ast H_{\Theta^\ast} = I - T_\Theta T_\Theta^\ast \text{.} \]
    Hence, by lemma \ref{lem:prop_36} and Douglas' lemma
    \[ \mathcal{H}(\Theta) = \operatorname{ran}(H_{\Theta^\ast}^\ast H_{\Theta^\ast})^{1/2} = \operatorname{ran} (H_{\Theta^\ast}^\ast) = \operatorname{ran} (H_{\breve{\Theta}})\]
    where the last step holds by 1. in lemma \ref{lem:th_identities}.
    \end{proof}
\end{lemma}

This gives the following strong consequence.

\begin{theorem}[Equivalence of model spaces]
    \label{thm:eq_of_model_spaces}
    Let $\Theta \in H^\infty(\mathbb{D}, \mathcal{B}(\mathcal{U}, \mathcal{Y}))$ be a left inner function and let $\Theta_\text{BLH}^c \in H(\mathbb{T}, \mathcal{B}(\mathcal{V}, \mathcal{Y}))$ be the BLH complementary factor of $\Theta$. Then 
    \[ \mathcal{H}(\begin{pmatrix} \Theta & \Theta^c_{\text{BLH}}\end{pmatrix}) \cong \mathcal{H}(\Theta^\sharp) \text{.} \]
    \begin{proof}
        By lemma \ref{lem:curto_4.5} we know that for the BLH complementary factor, $\Theta^c_{\text{BLH}} \in H^\infty(\mathbb{T}, \mathcal{B}(\mathcal{V}, \mathcal{Y}))$, we have 
        \[ \operatorname{ker} H_{\Theta^\ast} = \Delta H^2(\mathbb{T}, \mathcal{U} \oplus \mathcal{V}) \]
        with $\Delta := \begin{pmatrix} \Theta & \Theta^c_{\text{BLH}} \end{pmatrix}$. Since $\Delta$ is left inner, we find 
        \[ (\operatorname{ker} H_{\Theta^\ast})^\perp = \mathcal{H}(\Delta)  \text{.} \]
        Secondly, choosing the right inner $\Theta^\sharp$ in lemma \ref{lem:mod_space} makes that
        \[ \operatorname{ran}(H_{\Theta^\ast}) = \mathcal{H}(\Theta^\sharp) \text{.} \]
        Hence,
        \[ W: \mathcal{H}(\Delta) \rightarrow \mathcal{H}(\Theta^\sharp): W = H_{\Theta^\ast}|_{\mathcal{H}(\Delta)} \]
        is a bijective map. Identify $\mathcal{H}(\Theta^\sharp)$ as the range space corresponding to $I - T_{\Theta^\sharp} T_{\Theta^\sharp}^\ast$ (see lemma \ref{lem:prop_36}). Then, let $H_{\Theta^\ast} = (H_{\Theta^\ast} H_{\Theta^\ast}^\ast)^{1/2} V$ be the polar decomposition of $H_{\Theta^\ast}$ for some partial isometry $V$ with initial space $(\operatorname{ker} H_{\Theta^\ast})^\perp = \mathcal{H}(\Delta)$ and final space 
        \[ \operatorname{clos \ ran} H_{\Theta^\ast} = \operatorname{clos} \mathcal{H}(\Theta^\sharp) = \operatorname{clos \ ran} (I - T_{\Theta^\sharp} T_{\Theta^\sharp}^\ast)^{1/2} = ( \operatorname{ker} (I - T_{\Theta^\sharp} T_{\Theta^\sharp}^\ast)^{1/2} )^\perp \]  
        (for details, see Theorem 6.10 in \cite{Reed1980} or Lemma 1.1.12 in \cite{Nikolski2002_2}). We find that for all $f \in \mathcal{H}(\Delta)$
        \[ \begin{split}\| W f \|_{\mathcal{H}(\Theta^\sharp)} &= \| (H_{\Theta^\ast} H_{\Theta^\ast}^\ast)^{1/2} V f \|_{\mathcal{H}(\Theta^\sharp)} = \| (I - T_{\Theta^\sharp} T_{\Theta^\sharp}^\ast)^{1/2} V f \|_{\mathcal{H}(\Theta^\sharp)} \\ &= \| V f \|_{H^2(\mathbb{T}, \mathcal{U})} = \| f \|_{H^2(\mathbb{T}, \mathcal{Y})} = \| f \|_{\mathcal{H}(\Delta)} \end{split} \]
        and thus $H_{\Theta^\ast}|_{\mathcal{H}(\Delta)}$ is unitary.
    \end{proof}
\end{theorem}

\begin{corollary}
    \label{cor:intertwiner}
    Let $\Theta \in H^\infty(\mathbb{D}, \mathcal{B}(\mathcal{U}, \mathcal{Y}))$ be a left inner function and let $\Delta := \begin{pmatrix} \Theta & \Theta^c_\text{BLH} \end{pmatrix}$. Then $H_{\Theta^\ast}|_{\mathcal{H}(\Delta)}$ unitarily intertwines the backward shift operator, $R_0$, on $\mathcal{H}(\Theta^\sharp)$ with the compressed shift operator, $C_0 := P_{\mathcal{H}(\Delta)} \mathbf{S}_{\mathcal{Y}}|_{\mathcal{H}(\Delta)}$, on $\mathcal{H}(\Delta)$.    
    \begin{proof}
        By Theorem \ref{thm:eq_of_model_spaces}
        \[ W: \mathcal{H}(\Delta) \rightarrow \mathcal{H}(\Theta^\sharp): W = H_{\Theta^\ast}|_{\mathcal{H}(\Delta)}\]
        is unitary. Moreover by lemma \ref{lem:commutator_of_shift}, we have that for all $f \in \mathcal{H}(\Delta)$
        \[ W C_0 f = H_{\Theta^\ast} P_{\mathcal{H}(\Delta)} \mathbf{S}_{\mathcal{Y}} f =  H_{\Theta^\ast} \mathbf{S}_{\mathcal{Y}} f  = \mathbf{S}_{\mathcal{U}}^\ast H_{\Theta^\ast} f = \mathbf{S}_{\mathcal{U}}^\ast |_{\mathcal{H}(\Theta^\sharp)} H_{\Theta^\ast} f = R_0 W f \]
        hence $W$ is an intertwiner.
    \end{proof}
    
\end{corollary}

\begin{definition}[Julia embedding, adapted from \cite{Lilleberg2019}] Let $\Sigma: \mathcal{H} \oplus \mathcal{U} \rightarrow  \mathcal{H} \oplus \mathcal{Y}$ be a passive realization. Its Julia embedding is given by the unitary
\[ \widehat{\Sigma}: \mathcal{H} \oplus \mathcal{U} \oplus \mathfrak{D}_{\Sigma^\ast} \rightarrow \mathcal{H} \oplus \mathcal{Y} \oplus \mathfrak{D}_\Sigma: \ \widehat{\Sigma} := \begin{pmatrix}
    \Sigma & D_{\Sigma^\ast}|_{\mathfrak{D}_{\Sigma^\ast}} \\
    D_{\Sigma} & - \Sigma^\ast|_{\mathfrak{D}_{\Sigma^\ast}}
\end{pmatrix}\]
where $D_T = (I - T^\ast T)^{1/2}$ is the defect operator and $\mathfrak{D}_T = \operatorname{clos \ ran} D_T$ is the defect space corresponding to $T$ (see also lemma 1.1.13 in \cite{Nikolski2002_2}). It can be conveniently regrouped as
\[ \widehat{\Sigma}: \mathcal{H} \oplus \widehat{\mathcal{U}} \rightarrow \mathcal{H} \oplus \widehat{\mathcal{Y}}: \ \widehat{\Sigma} := \begin{pmatrix}
    A & \widehat{B} \\
    \widehat{C} & \widehat{D}
\end{pmatrix}\]
with $\widehat{\mathcal{U}} = \mathcal{U} \oplus \mathfrak{D}_{\Sigma^\ast}$, $\widehat{\mathcal{Y}} = \mathcal{Y} \oplus \mathfrak{D}_\Sigma$ and
\[ \widehat{B} = \begin{pmatrix}
    B & B_1
\end{pmatrix}, \quad \widehat{C} = \begin{pmatrix}
    C \\ C_1
\end{pmatrix}, \quad \widehat{D} = \begin{pmatrix}
    D & D_{12} \\
    D_{21} & D_{22} 
\end{pmatrix} \text{.}\]
Moreover, its characteristic function is given by
\[ \theta_{\widehat{\Sigma}}(z) = \begin{pmatrix}
D + zC(I - zA)^{-1}B
&
D_{12} + zC(I - zA)^{-1}B_1
\\[6pt]
D_{21} + zC_1(I - zA)^{-1}B
&
D_{22} + zC_1(I - zA)^{-1}B_1 
\end{pmatrix} \text{.}
\]
\end{definition}

\begin{remark}
    Note that for a coisometric realization, $\Sigma$, the defect operator $D_{\Sigma^\ast} = 0$ and hence its Julia embedding is given by 
    $ \begin{pmatrix}
        \Sigma \\
        D_\Sigma
    \end{pmatrix} 
    $. 
\end{remark}

\begin{lemma}
    \label{lem:Julia_transfer}
    Let $\Sigma$ be a passive realization and let $\widehat{\Sigma}$ be its Julia embedding. Then if $\Sigma$ is observable (controllable, stable, costable) then $\widehat{\Sigma}$ is observable (controllable, stable, costable).
    \begin{proof}
        This easily follows from the fact that $B \mathcal{U} \subseteq \widehat{B} \widehat{\mathcal{U}}$ and $C^\ast \mathcal{Y} \subseteq \widehat{C}^\ast \widehat{\mathcal{Y}}$ and the fact that the main operator does not change.
    \end{proof}
\end{lemma}

\begin{theorem}
    \label{thm:eq_of_real}
    Let $\Theta \in H^\infty(\mathbb{D}, \mathcal{B}(\mathcal{U}, \mathcal{Y}))$ be a left inner function and let $U$ be the canonical realization of $\Theta^\sharp$. Let $\widehat{U}$ be the Julia embedding of $U$ and let $\Theta_\text{BLH}^c \in H(\mathbb{T}, \mathcal{B}(\mathcal{V}, \mathcal{Y}))$ be the BLH complementary factor of $\Theta$. Then there exists a constant unitary $Q: \mathfrak{D}_U \rightarrow \mathcal{V}$ such that
    \[ \theta_{\widehat{U}^\ast} = \begin{pmatrix} \Theta & \Theta^c_{\text{BLH}} Q \end{pmatrix} \text{.}\]
    Moreover, $\widehat{U}^\ast$ is a minimal realization.
    \begin{proof}
        Let $\widehat{V}$ be the canonical realization of $\begin{pmatrix} \Theta  & \Theta_\text{BLH}^c \end{pmatrix}$ and denote 
        \[ \begin{split} &\widehat{U}: \mathcal{H}(\Theta^\sharp) \oplus \mathcal{Y} \rightarrow \mathcal{H}(\Theta^\sharp) \oplus \mathcal{U} \oplus \mathfrak{D}_U: \quad \widehat{U} = \begin{pmatrix} A & B \\ C & D \\ C_1 & D_1 \end{pmatrix}, \\ &\widehat{V}: \mathcal{H}(\Delta) \oplus \mathcal{U} \oplus \mathcal{V} \rightarrow \mathcal{H}(\Delta) \oplus \mathcal{Y}: \quad \widehat{V} = \begin{pmatrix} E & F & F_1 \\ G & H & H_1 \end{pmatrix} \text{.} \end{split}  \]
        We first show that the unitary $W$ constructed in Theorem \ref{thm:eq_of_model_spaces} is such that 
        \[ A = W E^\ast W^\ast  \quad (1), \quad 
            B = W G^\ast  \quad (2), \quad
            C = F^\ast W^\ast  \quad (3), \quad
            D = H^\ast  \quad (4) \]
        and then
        \[ Q \begin{pmatrix} C_1 & D_1 \end{pmatrix} = \begin{pmatrix} F_1^\ast W^\ast & H_1^\ast \end{pmatrix} \]
        for some constant unitary $Q: \mathfrak{D}_U \rightarrow \mathcal{V}$. Before we do so, it can be verified that for all $f \in H^2(\mathbb{T}, \mathcal{Y})$
            \[ H_{\Theta^\ast} f = \sum_{k=0}^\infty R_0^{k+1} \Theta^\sharp f_k \]
        where $f_k$ are the Fourier components of $f$. Then
        \begin{enumerate}
            \item[(1)] We have $E = R_0$. Note that because $\Delta$ is left inner, $\mathcal{H}(\Delta)$ inherits the adjoint from the ambient $H^2(\mathbb{T}, \mathcal{Y})$, hence $E^\ast$ is equal to the compressed shift $C_0$. Hence, corollary \ref{cor:intertwiner} says that $A W = W E^\ast$.\footnote{For non inner $\Delta$ this need not hold and we would generally have $(R_0^\ast f)(z) = z f(z) - \Delta(z) \tau^\ast f$.}
            \item[(2)] We have $G = e(0)$ and it can be shown (see \cite{Ball2014}) that $e(0)^\ast = I - \Delta(z) \Delta^\ast(0)$. Hence, for all $y \in \mathcal{Y}$
            \[ W G^\ast y = W e(0)^\ast y = H_{\Theta^\ast}  y - H_{\Theta^\ast} \Delta(z) \Delta^\ast(0) y \text{.} \]
            By lemma \ref{lem:curto_4.5}, we know that $\Delta(z) \Delta^\ast(0) y \in \ker H_{\Theta^\ast}$ and hence
            \[ W G^\ast y = R_0 \Theta^\sharp y = \frac{\Theta^\sharp(z) - \Theta^\sharp(0)}{z}y = B y \text{.} \]
            \item[(3)] Consider that $C = e(0)$ and hence for all $f \in \mathcal{H}(\Delta)$ we have
            \[ C W f = e(0) \sum_{k=0}^\infty R_0^{k+1} \Theta^\sharp f_k = \sum_{k=0}^\infty \Theta_{k+1}^\ast f_k \text{.} \]
            At the same time we have
            \[ F = \frac{\Theta(z) - \Theta(0)}{z} \text{.}\]
            Since $\mathcal{H}(\Delta)$ inherits the $H^2(\mathbb{T}, \mathcal{Y})$ inner product, we have for all $g \in \mathcal{H}(\Delta)$ and $u \in \mathcal{U}$
            \[ \langle F u | g \rangle_{\mathcal{H}(\Delta)} = \left\langle u \bigg| \sum_{k=0}^\infty \Theta^\ast_{k+1} g_k \right\rangle_{\mathcal{U}} \]
            and hence $C = F^\ast W^\ast$.
        \item[(4)] We have $D = \Theta^\sharp(0) = \Theta^\ast(0) = H^\ast$.
        \end{enumerate}
        Hence we conclude that 
        \[ \begin{pmatrix}
        W & 0 & 0 \\
        0 & I & 0 \\
        0 & 0 & I
        \end{pmatrix} \widehat{V}^\ast  \begin{pmatrix}
        W^\ast & 0 \\
        0 & I
        \end{pmatrix} = \begin{pmatrix}
            U \\ R
        \end{pmatrix}, \quad \widehat{U} = \begin{pmatrix} U \\ P \end{pmatrix} \]
        with $R = \begin{pmatrix} F_1^\ast W^\ast & H_1^\ast \end{pmatrix}$ and $P = \begin{pmatrix}C_1 & D_1 \end{pmatrix}$. Suppose next that for $w \in \mathcal{U} \oplus \mathcal{V}$ we have $\Delta w \in \mathcal{H}(\Delta)$, then $w = 0$ since $w \in H^2(\mathbb{D}, \mathcal{U} \oplus \mathcal{V})$ and $\Delta H^2(\mathbb{D}, \mathcal{U} \oplus \mathcal{V}) \perp \mathcal{H}(\Delta)$. Then Theorem 3.12 (1.) in \cite{Ball2014} implies that $\widehat{V}$ and hence $\widehat{V}^\ast$ must be conservative. Then using that both $\widehat{V}^\ast$ and $\widehat{U}$ are isometric gives $R^\ast R = P^\ast P$, i.e. $R = Q P$ for some partial isometry $Q : \mathfrak{D}_U \rightarrow \mathcal{V}$ with initial space $\operatorname{clos \ ran} P$ and final space $\operatorname{clos \ ran} R$ by Douglas' lemma. Moreover using that both $\widehat{V}^\ast$ and $\widehat{U}$ are coisometric gives $I = Q Q^\ast$ and $I = Q^\ast Q$, i.e. $Q$ is unitary. Hence, we also find that 
        \[ \theta_{\widehat{U}^\ast} = \begin{pmatrix} \Theta & \Theta^c_{\text{BLH}} Q \end{pmatrix} \text{.}\]
        Lastly, we know that $\widehat{V}$ is observable and thus $\widehat{V}^\ast$ is controllable. By identification, this means that $\widehat{U}$ is controllable and hence $\widehat{U}^\ast$ is observable. Moreover, $\widehat{U}$ is the Julia embedding of observable $U$, which by lemma \ref{lem:Julia_transfer} means that $\widehat{U}$ is observable and thus $\widehat{U}^\ast$ is controllable. Hence $\widehat{U}^\ast$ is minimal.
    \end{proof}

\end{theorem}

Now we are equiped to relate the completability of $\Theta$ to the properties of the realizations. To do so, we need the additional notion of strong stability.

\begin{definition}[Strongly stable, strongly costable, strongly bistable]
    A realization $\Sigma: \mathcal{H} \oplus \mathcal{U} \rightarrow \mathcal{H} \oplus \mathcal{Y}$ of $F \in H^\infty(\mathbb{D},\mathcal{B}(\mathcal{U}, \mathcal{Y}))$ with main operator $A$ is said to be strongly stable (costable) if for all $x \in \mathcal{H}$ we have $\| A^n x\| \rightarrow 0$ ($\| (A^\ast)^n x\| \rightarrow 0$) as $n \rightarrow \infty$. $\Sigma$ is said to be strongly bistable if it is both strongly stable and strongly costable.
\end{definition}

\begin{lemma}[Adapted from Theorem 4.2 in \cite{Lilleberg2019}]
    \label{lem:Lilleberg}
    Let $\Sigma$ be a simple conservative realization of $F \in H^\infty(\mathbb{D}, \mathcal{B}(\mathcal{U}, \mathcal{Y}))$ with main operator $A$. Then
    \[ A \text{ is strongly stable} \iff F \text{ is left inner}, \quad A \text{ is strongly costable} \iff F \text{ is right inner} \text{.}\]
\end{lemma}

\begin{theorem}
    \label{thm:completability_and_main_operator}
    Let $\Theta \in H^\infty(\mathbb{D}, \mathcal{B}(\mathcal{U}, \mathcal{Y}))$ be a left inner function and let $U$ be the canonical realization of $\Theta^\sharp$ with main operator $A$. Then
    \[ A \text{ is strongly stable} \iff \Theta \text{ is completable} \text{.} \]
    \begin{proof}
        Let $\widehat{U}$ be the canonical realization of $\Delta := \begin{pmatrix} \Theta & \Theta^c_{\text{BLH}} \end{pmatrix}$ with main operator $\widehat{A}$. We have seen in Theorem \ref{thm:eq_of_real} that $\widehat{A}^\ast$ is unitarily equivalent to $A$. Moreover note that $\widehat{U}$ is observable by lemma \ref{lem:Ball}, hence simple by remark \ref{rem:simple} and is moreover conservative as shown in Theorem \ref{thm:eq_of_real}. Then by interwining, lemma \ref{lem:Lilleberg}, lemma \ref{lem:curto_4.20} and \ref{lem:curto_4.21} we have
        \[ \begin{split} &A \text{ is strongly stable} \iff \widehat{A} \text{ is strongly costable} \\ \iff &\Delta \text{ is right inner} \iff \Theta \text{ is completable} \end{split} \]
    \end{proof}
\end{theorem}

This has a direct implication that ties back to the original motivation.

\begin{corollary}
    \label{cor:completability}
    Let $\varphi \in L^\infty(\mathbb{T} \times \mathbb{T})$ be a unital Schur-class kernel. Let $\mathcal{K}_{\widetilde{\varphi}}$ be the reproducing kernel Hilbert space with kernel
    \[ K(z,w) = \frac{1 - \widetilde{\varphi}(\overline{z},\overline{w})}{1 - z \overline{w}} \text{.} \]
    Then the minimal Stinespring vector of $\varphi$ is completable to a two-sided inner function if and only if the backward shift on $\mathcal{K}_{\widetilde{\varphi}}$ is strongly stable.
    \begin{proof}
        Let $[\widetilde{v}]$ be the essentially unique minimal Stinespring vector (see corollary \ref{cor:Stinespring_rep}), which is left inner by corollary \ref{cor:unital_real_schur_mult}. Then the de Branges Rovnyak space belonging to $[\widetilde{v}]^\sharp$ is generated by the kernel
        \[ \frac{1 - [\widetilde{v}]^\sharp(z) \breve{[\widetilde{v}]}(w) }{1 - z \overline{w}}  = \frac{1 -\widetilde{\varphi}(\overline{z},\overline{w})}{1 - z \overline{w}} = K(z,w) \text{.} \]
        Then Theorem \ref{thm:completability_and_main_operator} implies that the backward shift operator on $\mathcal{K}_{\widetilde{\varphi}}$ is strongly stable if and only if $[\widetilde{v}]$ is completable.
    \end{proof}
\end{corollary}

We now discuss some results that allow us to compute a complementary factor in finite dimensions and rational inner functions.

\begin{lemma}[Adapted from lemma 3.3 in \cite{Frazho2018}]
    \label{lem:Frazho_3_3}
    Let $\Theta$ be a right inner function in $H^\infty(\mathbb{D}, \mathcal{B}(\mathcal{U}, \mathcal{Y}))$ where $\mathcal{U}$ and $\mathcal{Y}$ are finite dimensional. Then
    \begin{enumerate}
    \small
        \item $\mathcal{H}(\Theta)$ is finite dimensional iff $\Theta$ is rational.
        \item The dimension of $\mathcal{H}(\Theta)$ equals the McMillan degree of $\Theta$.
    \end{enumerate}
    \begin{proof}
        We know that $\mathcal{H}(\Theta) = \operatorname{ran} H_{\breve{\Theta}}$ by lemma \ref{lem:mod_space}. The range of a Hankel operator $H_{\breve{F}}$ is finite dimensional if and only if its symbol $F$ is rational. Moreover, the dimension of the range of the Hankel operator $H_{\breve{F}}$ equals the McMillan degree of $F$.
    \end{proof}
\end{lemma}

\begin{lemma}[Adapted from Theorem 3.5 in \cite{Fritzsche2008}]
    \label{thm:fritzsche_3_5}
    Let $\Sigma$ be a finite-dimensional, conservative realization and $\theta_\Sigma$ be its characteristic function. Then $\theta_\Sigma$ is a two-sided, rational inner function.
\end{lemma}

\begin{theorem}[Finite dimensional completions]
    \label{thm:comp_facts}
    Let $\Theta$ be a rational, left inner function in $H^\infty(\mathbb{T}, \mathcal{B}(\mathcal{U}, \mathcal{Y}))$, where
    $\dim \mathcal{U},\dim \mathcal{Y}<\infty$. Let $U$ denote the canonical realization of $\Theta^\sharp$ and let $\widehat{U}$ be its Julia embedding. Then $\widehat{U}$ is such that $\theta_{\widehat{U}^\ast}$ is rational and the minimal two-sided inner completion of $\Theta$.
    \begin{proof}
        By rationality of $\Theta^\sharp$ and the fact that $\dim \mathcal{U} \leq \dim \mathcal{Y} < \infty$, lemma \ref{lem:Frazho_3_3} implies that the de Branges-Rovnyak space
        $\mathcal{H}(\Theta^\sharp)$ is finite dimensional. Hence the canonical realization
        $U$ is a finite-dimensional coisometric realization. The Julia embedding is equal to 
        \[ \widehat{U} = \begin{pmatrix} U \\ W^\ast \end{pmatrix} \]
        where $W$ is such that its columns form an orthonormal basis for $\ker U$. Then, lemma \ref{thm:fritzsche_3_5} implies that the characteristic function of the adjoint $\theta_{\widehat{U}^\ast}$ is rational, two-sided inner and by Theorem \ref{thm:eq_of_real} equal to
        \[ \theta_{\widehat{U}^\ast} = \begin{pmatrix} \Theta & \Theta^c_\text{BLH} Q \end{pmatrix} \]
        for some constant unitary $Q$. By definition $\Theta^c_\text{BLH} Q$ is minimal.
    \end{proof}
\end{theorem}
\section{Application of results}
\label{sec:application}


In order to show an application of our theory, we provide a physically meaningfull example. In particular, we start with a TIC quantum channel and explicitly derive a minimal system that implements it. Along the way, we will use the algorithms developed in appendix \ref{ap:Algorithms}. The algorithms mostly follow from the theory in the previous chapters.

\begin{example}
    Suppose that an initially unknown SISO system is described by the Schur multiplier $M_\varphi$ with symbol $\varphi: \mathbb{R} \times \mathbb{R} \rightarrow \mathbb{C}$ given by
    \[ \varphi(\omega, \omega') = \frac{1+8\sin(2\pi \omega/\omega_0)\sin(2\pi \omega'/\omega_0)}{(\cos(2\pi \omega/\omega_0) + 3 i \sin(2\pi \omega/\omega_0)) (\cos(2\pi \omega'/\omega_0) - 3 i \sin(2\pi \omega'/\omega_0))} \]
    where $\omega, \omega'$ denote frequency and $\omega_0 > 0$ is some characteristic frequency. 
    This function can be written as the composition $\varphi = \eta \circ (\zeta, \zeta)$ of the kernel $\eta: \mathbb{T} \times \mathbb{T} \rightarrow \mathbb{C}$ and the (singular) inner function $\zeta:\mathbb{R} \rightarrow \mathbb{T}$ where
    \[ \eta(x,y) = \frac{2 (\overline{x}^2 - 1)(y^2 - 1) + \overline{x} y}{\left(\overline{x}^2 - 2 \right) \left(y^2 - 2 \right)}, \qquad \zeta(\omega) = \exp(2\pi i \omega/ \omega_0) \text{.}\]
    It is readily verified that $\eta$ satisfies the properties of unital Schur-class kernel. This can be seen from the form of $\eta$ which immediately gives $\dim H_{\tilde{\eta}} \leq 2 $ by remark \ref{rem:Papadakis} and the fact that $\eta(x,x) = 1$. Indeed, picking the two points $z_1= 1/2, z_2 = -1/2$ we find that $\eta(\cdot, z_1), \eta(\cdot, z_2)$ are independent RKHS vectors  which can be orthogonalized using algorithm \ref{alg:RKHS_basis}. This results in the left inner function
    \[ \Theta: \mathbb{T} \rightarrow \mathcal{B}(\mathbb{C}, \mathbb{C}^2): x \mapsto \Theta(x) := \frac{1}{\sqrt{11} (x^2 - 2)}\begin{pmatrix}
        \sqrt{2} (3 x^2 - x - 3) \\ 2 x^2 + 3 x - 2
    \end{pmatrix}\]
    which induces a Stinespring representation, with $\eta(x,y) = \Theta^\ast(x) \Theta(y)$. Using algorithm \ref{alg:complementary_factor}, we find the completed two-sided inner function to be
    \[ \Delta' : \mathbb{T} \rightarrow \mathcal{B}(\mathbb{C}^2, \mathbb{C}^2): x \mapsto \Delta'(x) := \frac{1}{\sqrt{11} (x^2 - 2 )} \begin{pmatrix}
        \sqrt{2} (3 x^2 - x - 3) & - 2 x^2 + 3 x + 2 \\ 2 x^2 + 3 x - 2 & \sqrt{2} (3 x^2 + x - 3)
    \end{pmatrix} \text{.} \]
    This result is minimal and essentially unique and we can use algorithm \ref{alg:blaschke_potapov} to get a unitarily equivalent, simpler expression. Now, the Blaschke-Potapov zeros and projectors are equal to $z = \pm 1/\sqrt{2}$ and
    \[ P = \frac{1}{2} \begin{pmatrix}
        1 & \pm 1 \\ \pm 1 & 1
    \end{pmatrix} \text{.} \]
    This yields the expression
    \[ \Delta: \mathbb{T} \rightarrow \mathcal{B}(\mathbb{C}^2, \mathbb{C}^2): x \mapsto \Delta(x) := \frac{1}{x^2 - 2} \begin{pmatrix}
        \sqrt{2} (x^2 - 1) & x \\ x & \sqrt{2} (x^2 - 1) 
    \end{pmatrix} \]
    which realizes $\varphi$ in the sense that $\Delta \circ \zeta$ is a two-sided inner function with
    \[ \varphi(\omega,\omega') = \operatorname{tr}\left[\Delta(\zeta(\omega')) \begin{pmatrix}
        1 & 0 \\
        0 & 0
    \end{pmatrix} \Delta^\ast(\zeta(\omega)) \right] \text{.} \]
    As it turns out, $\Delta \circ \zeta$ is precisely the scattering matrix of a ring resonator that is (symmetrically) coupled to two waveguides. The choice of which output port is regarded is "forgotten", which is mathematically represented by the trace. For details on such resonator structures, we refer to \cite{Rabus2007}, chapter 2.

    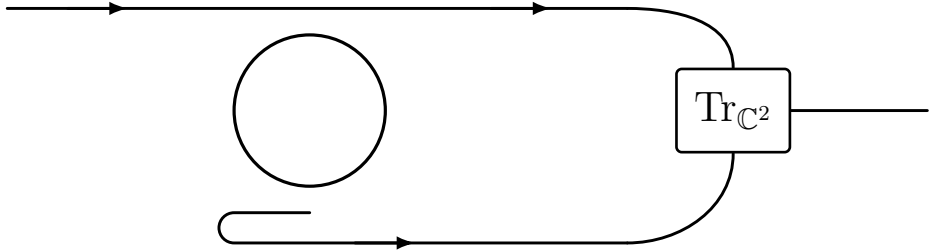
\begin{figure}[!tbh]
        \centering
        \begin{tikzpicture}[
                line cap=round,
                line join=round,
                >=Latex
            ]

            \def\R{1.0}
            \def\gap{0.35}
            \def\rb{0.2} 

            \draw[resonator] (0,0) circle (\R);

            \draw[waveguide] (-4,\R+\gap) -- (4.2,\R+\gap);

            \draw[waveguide] (0.0,-\R-\gap) -- (-1.0,-\R-\gap)
            arc[start angle=90,end angle=270,radius=\rb]
            -- (4.2,{-\R-\gap-2*\rb});

            \draw[quantum] (-3.2,\R+\gap) -- (-2.4,\R+\gap);
            \draw[quantum] (2.4,\R+\gap) -- (3.2,\R+\gap);

            \draw[quantum] (0.6,{-\R-\gap-2*\rb}) -- (1.4,{-\R-\gap-2*\rb});

            \node[tracebox, font=\Large] (T) at (5.6,0) {$\mathrm{Tr}_{\mathbb{C}^2}$};

            \draw[waveguide] (4.2,\R+\gap) to[out=0,in=90] (T.north);
            \draw[waveguide] (4.2,{-\R-\gap-2*\rb}) to[out=0,in=-90] (T.south);

            \draw[waveguide] (T.east) -- ++(1.8,0);

        \end{tikzpicture}
        \caption{A single ring attached to two waveguides. The output waveguide is traced out, resulting in a mixed state output.}
    \end{figure}
\end{example}
\section{Conclusion and discussion}
We have seen that both questions posed in the introduction are resolved; It is indeed always possible to view Schur-class kernels, $\varphi$, as a single causal system through its factorization. Its natural Stinespring dilation vector $[v]$ is minimal and therefore essentially unique. We have moreover seen that the dilation vector of a TIC quantum channel, $[v]$, is left inner and can be completed to a two-sided inner function whenever the main operator of the de Branges-Rovnyak realization of $[v]^\sharp$ is strongly stable. The adjoint of the Julia embedding of such a realization then gives a minimal two-sided inner completion $\begin{pmatrix} [v] & [v]^c_\text{BLH} Q \end{pmatrix}$ up to constant unitary $Q$. For finite dimensions and rational $[v]$ the minimal completion is computable.

An example of a TIC quantum channel that we cannot easily algorithmically realize as a causal system is 
\[ \varphi(x,y) = (1 + \overline{s(x)} s(y))/2 \]
with $s: \mathbb{T} \rightarrow \mathbb{T}$ some singular inner function. This has to do with the fact that the model space is not of finite dimension (cf. lemma \ref{lem:Frazho_3_3}), such that we cannot straightforwardly complete the coisometric realization to a conservative one. What is especially unsatisfying, is that manual inspection readily gives a solution
\[ U(x) = \frac{1}{\sqrt{2}}\begin{pmatrix}
    1 & 1 \\
    s(x) & -s(x)
\end{pmatrix} \text{.} \] 
This brings us to the first follow-up question:
\begin{questionbox}
    Can we algorithmically find two-sided inner completions of singular left inner functions?
\end{questionbox}
As discussed before, the second limitation is related to the reliance of our computational approach on finite-dimensional left inner functions. We thus pose the second follow-up question:
\begin{questionbox}
    Can two-sided inner completions of infinite-dimensional completable left inner functions be computed algorithmically?
\end{questionbox}
An example of such a situation is given by a modified Szeg\H{o} kernel
\[ \varphi(x,y) = \frac{1-|a|^2}{1 - |a|^2 \ \overline{x} y }\]
with $a \in \mathbb{D} \setminus \{0\}$, which admits the minimal factorization $\varphi(x,y) = \langle v (x) | v(y) \rangle_{\ell^2}$ with 
$ v_n(x) = \sqrt{1-|a|^2} \ (a x)^{n} $
for $n \in \mathbb{N}_0$.

Lastly, we note that the developed SISO theory of Schur-class kernels can quite straightforwardly be generalized to MIMO systems. To do so, future work should include some adapted notions, such as measurable \textit{operator-valued} Schur multipliers, and adaptation of the algorithms for operator-valued RKHS function spaces.

\backmatter

\bmhead{Acknowledgments}
We express our gratitude towards Joseph Ball, Georg Prokert and André Ran for valuable conversations. We moreover acknowledge OpenAI's ChatGPT for assisting with the identification of relevant sources and for discussing mathematical ideas. We bear full responsibility for any errors in the work.

\section*{Statements and Declarations}

\subsection*{Funding}
Partial financial support has been received from the Challenges in
CyberSecurity Zwaartekracht program and the Groeifonds project Quantum Delta NL KAT-2.

\subsection*{Competing interests}
The authors have no competing interests to declare that are relevant to the content of this article.

\subsection*{Data availability}
This article does not include any data.


\begin{appendices}
\begin{appendix}
    \section{Proof of lemma \ref{lem:Schur_class}}
    \label{ap:proof_of_lemma14}
    Note first that $\cdot^\ast: \mathcal{B}(\mathcal{U}, \mathcal{Y}) \rightarrow \mathcal{B}(\mathcal{Y}, \mathcal{U}): A \mapsto A^\ast$ is a conjugate linear isometric bijection. Thus for left inner $\widetilde{\Theta} \in H^\infty(\mathbb{D}, \mathcal{B}(\mathcal{U}, \mathcal{Y}))$ the following limit holds
    \[ \left\| \frac{\widetilde{\Theta}^\sharp(z_0 + h) - \widetilde{\Theta}^\sharp(z_0)}{h} - \widetilde{\Theta}'(\overline{z_0})^\ast \right\| = \left\| \frac{\widetilde{\Theta}(\overline{z_0} + \overline{h}) - \widetilde{\Theta}(\overline{z_0})}{\overline{h}} - \widetilde{\Theta}'(\overline{z_0}) \right\| \rightarrow 0  \]
    as $h \rightarrow 0$ for all $z_0 \in \mathbb{D}$ and hence $\widetilde{\Theta}^\sharp$ is strongly analytic. Moreover, for any $u \in \mathcal{U}$ with $\| u \| = 1$ and $z \in \mathbb{D}$ we have
    \[ \begin{split} \| \widetilde{\Theta}(z) u \| &= \sup_{y \in \mathcal{Y}, \|y \| = 1} | \langle y | \widetilde{\Theta}(z) u \rangle | \leq  \sup_{y \in \mathcal{Y}, \|y \| = 1} \underset{x \in \mathbb{T}}{\operatorname{essup}} \ |\langle y | \Theta(x) u  \rangle| \\ &\leq \ \sup_{y \in \mathcal{Y}, \|y \| = 1} \underset{x \in \mathbb{T}}{\operatorname{essup}} \ \| y \| \| \Theta(x) u  \| = 1 \end{split} \]
    by the maximum principle of V.I. Smirnov \cite{Katsnelson2007}, because $\langle y | \widetilde{\Theta}(z) u \rangle$ is an $H^\infty(\mathbb{D})$ function and the boundary function of $\widetilde{\Theta}$ is an isometry. For the reflection, we have for any $y \in \mathcal{Y}$ such that $\|y\| = 1$ and $z \in \mathbb{D}$ that
    \[ \begin{split} &\| \widetilde{\Theta}^\sharp(z) y \| = \sup_{u \in \mathcal{U}, \|u \| = 1} | \langle u | \widetilde{\Theta}^\sharp(z) y \rangle | = \sup_{u \in \mathcal{U}, \|u \| = 1} | \langle \widetilde{\Theta}(\overline{z}) u | y \rangle |  \\ \leq &\sup_{u \in \mathcal{U}, \|u \| = 1} \underset{x \in \mathbb{T}}{\operatorname{essup}} \ | \langle \Theta(\overline{x}) u | y \rangle | \leq 1 \end{split} \]
    because $\langle \widetilde{\Theta}(\overline{z}) u | y \rangle$ is an $H^\infty(\mathbb{D})$ function.

    \section{Proof of Theorem \ref{thm:1}}
    \label{ap:proof_of_thm1}
        Let $\{ e_k \}_{k=1}^d \subset \mathcal{U}$ be an o.n.b. of $\text{ran}(P)$ with $d = \dim \text{ran}(P) < \infty$ and let $\{ h_\ell \}_{\ell = 1}^{\dim \mathcal{Y}}$ be an o.n.b. of $\mathcal{Y}$. First we prove boundedness. We have for all $z \in \mathbb{D}$
        \[ \| \widetilde{S}_\scriptr(z) P \|_\text{HS}^2 = \sum_k \| \widetilde{S}_\scriptr(z) e_k \|_{\mathcal{Y}}^2 \leq  \sum_k \| e_k \|_{\mathcal{U}}^2 = d < \infty \]
        making $\widetilde{S}_\scriptr(z) P$ Hilbert-Schmidt. Define $G: R \rightarrow \mathcal{S}(\mathbb{D},\mathcal{B}(\text{ran}(P), \mathcal{Y})):  \scriptr \mapsto \widetilde{G}_\scriptr := \widetilde{S}_\scriptr P$ and $G(z) : \scriptr \mapsto \widetilde{G}_\scriptr(z) := \widetilde{S}_\scriptr(z) P$ for each $z \in \mathbb{D}$. Observe that for $z,w \in \mathbb{D}$ we have that $\widetilde{G}_\scriptr(z)^\ast \widetilde{G}_\scriptr(w)$ is trace class and
        \[ \begin{split} Tr[\widetilde{G}_\scriptr(z)^\ast \widetilde{G}_\scriptr(w)] = &\sum_{k = 1}^d \langle e_k | \widetilde{G}_\scriptr(z)^\ast \widetilde{G}_\scriptr(w) | e_k \rangle = \sum_{\ell = 1}^{\dim \mathcal{Y}} \sum_{k = 1}^d \langle e_k | \widetilde{G}_\scriptr(z)^\ast | h_\ell \rangle \langle h_\ell | \widetilde{G}_\scriptr(w) | e_k \rangle \\ =  &\sum_{\ell = 1}^{\dim \mathcal{Y}} \langle h_\ell | \widetilde{G}_\scriptr(w) \widetilde{G}_\scriptr(z)^\ast | h_\ell \rangle = Tr[\widetilde{G}_\scriptr(w)  \widetilde{G}_\scriptr(z)^\ast]  \quad (\star) \end{split} \]
        so $\widetilde{S}_\scriptr(w) \ P \ \widetilde{S}_\scriptr(z)^\ast$ is trace-class and we may define $\widetilde{\varphi}_\scriptr(z,w) := \operatorname{tr}[ \widetilde{S}_\scriptr(w) \ P \ \widetilde{S}_\scriptr(z)^\ast ]$. Since $\scriptr \mapsto \widetilde{G}_\scriptr(z)$ is WOT measurable -- in fact strongly measurable by lemma \ref{lem:Dunford} and \ref{lem:Ende} because of the finite dimensional range of $P$ -- and ($\star$) is a countable sum of measurable maps, $\scriptr \mapsto \widetilde{\varphi}_\scriptr(z,w)$ is measurable for all $z,w \in \mathbb{D}$. Then for all $z,w \in \mathbb{D}$ we have that $|\widetilde{\varphi}_\scriptr(z,w) | \leq d < \infty$  and thus $\int_R | \widetilde{\varphi}_\scriptr | \ d\mu(\scriptr) \leq d \ \mu(R) < \infty$ because $\mu$ is finite, which means that the map $\scriptr \mapsto \widetilde{\varphi}_\scriptr$ is indeed integrable; denote $\widetilde{\varphi} = \int_R \widetilde{\varphi}_\scriptr \ d\mu(\scriptr)$. 
        
        Next, by the fact that $|\widetilde{\varphi}_\scriptr(z,w) | \leq d < \infty$ for $z,w \in \mathbb{D}$ the dominated convergence Theorem implies that $\widetilde{\varphi}(z,w)$ is separately continuous. Then by Tonelli-Fubini
        \[ \int_\gamma \widetilde{\varphi}(\overline{z},w_0) \ dz = \int_\gamma \int_R \widetilde{\varphi}_\scriptr(\overline{z},w_0) \ d\mu(\scriptr) \ dz = \int_R \int_\gamma \widetilde{\varphi}_\scriptr(\overline{z},w_0) \ dz  \ d\mu(\scriptr)  = 0 \]
        for all fixed $w_0 \in \mathbb{D}$ and closed curves $\gamma \subset \mathbb{D}$ and 
        \[ \int_\gamma \widetilde{\varphi}(z_0,w) \ dw = \int_\gamma \int_R \widetilde{\varphi}_\scriptr(z_0,w) \ d\mu(\scriptr) \ dw = \int_R \int_\gamma \widetilde{\varphi}_\scriptr(z_0,w) \ dw  \ d\mu(\scriptr)  = 0 \]
        for all fixed $z_0 \in \mathbb{D}$ and closed curves $\gamma \subset \mathbb{D}$. Hence, $\widetilde{\varphi}$ is anti-analytic in the first variable, and analytic in the second by Morera's Theorem, which means that it is sesquianalytic.
        
        Then to prove positive semidefiniteness, note that $\widetilde{\varphi}$ is Hermitian and let $\{z_n\}_{n = 1}^N \subset \mathbb{D}$ and $\{\alpha_n\}_{n = 1}^N \subset \mathbb{C}$ for some $N \in \mathbb{N}$. We have
        \[ \begin{split} \sum_{n,m = 1}^N \overline{\alpha_n} \alpha_m \widetilde{\varphi}(z_n, z_m) = &\int_R \left\langle \left( \sum_{n = 1}^N  \alpha_n \widetilde{G}_\scriptr(z_n) \right) \ \bigg| \ \left( \sum_{m = 1}^N \alpha_m \widetilde{G}_\scriptr(z_m) \right) \right\rangle_\text{HS} \ d\mu(\scriptr) \\ = &\int_R \left\| \sum_{n = 1}^N  \alpha_n \widetilde{G}_\scriptr(z_n)   \right\|_\text{HS}^2 \ d\mu(\scriptr) \geq 0 \end{split}\]
        by linearity of the trace and the non-negativity of the Hilbert-Schmidt norm.

        Lastly, by lemma \ref{lem:Nikolski_adapted} the limit $G_\scriptr(x) = \lim_{z \rightarrow x} \widetilde{G}_\scriptr(z)$ exists strongly (and hence in SOT) non-tangentially a.e. for $x \in \mathbb{T}$ and $x \mapsto G_\scriptr(x)$ is bounded and strongly measurable for all $\scriptr \in R$. Define the "bad set"
        \[ N_G := \{ (\scriptr, x) \in R \times \mathbb{T} \ | \ \widetilde{G}_\scriptr \text{ does not have a SOT non-tangential limit at } x \} \]
        and note that $\widetilde{G}_\scriptr \text{ does not have a SOT non-tangential limit at } x$ means that
        \[ \begin{split} \exists_{u \in \text{ran}(P), \alpha > 1, \varepsilon > 0} \ \forall_{\delta > 0}  \ \exists_{z,w \in \Gamma_\alpha(x), |z - x|, |w - x| < \delta}: \|\widetilde{G}_\scriptr( z ) u - \widetilde{G}_\scriptr( w ) u \| \geq \varepsilon \end{split}\text{.} \]
        For this, it is equivalent that  
        \[ \begin{split} \exists_{v \in D_{\text{ran}(P)},\beta \in \mathbb{N}_{>1}, k \in \mathbb{N}} \ \forall_{n \in \mathbb{N}} \ \exists_{\eta,\zeta \in D_{\mathbb{D}} \cap \Gamma_\alpha(x),  |\eta - x|, |\zeta - x| < \frac{1}{n}}: \|\widetilde{G}_\scriptr( \eta ) v - \widetilde{G}_\scriptr( \zeta ) v \| > \frac{1}{k} \end{split} \]
        for $D_{\text{ran}(P)}$ some countable dense set of $\text{ran}(P)$ and $D_{\mathbb{D}}$ some countable dense set of $\mathbb{D}$. Indeed, let $u \in \text{ran}(P), \alpha > 1, \varepsilon > 0$ be such a tuple. Let $k = \lceil 3/\varepsilon \rceil$, $\beta = \lceil \alpha \rceil$ and $v \in D_{\text{ran}(P)}$ such that $\| v - u \| <\varepsilon/6$. Then let $n \in \mathbb{N}$ be fixed. We know that there exists $z,w \in \Gamma_\alpha(x), |z - x|, |w - x| < 1/(2n)$ such that $\| \widetilde{G}_\scriptr( z ) u - \widetilde{G}_\scriptr( w ) u \| \geq \varepsilon$. Let $\eta, \zeta \in D_\mathbb{D} \cap \Gamma_\beta(x)$ be such that $|\eta - z|, |\zeta - w| < 1/(2n)$ and $\| \widetilde{G}_\scriptr(\eta) u - \widetilde{G}_\scriptr(z) u \|, \| \widetilde{G}_\scriptr(\zeta) u - \widetilde{G}_\scriptr(w) u \| < \varepsilon/6$ which is possible by density and continuity of $z \mapsto \widetilde{G}_\scriptr(z) u$. We find that $\eta, \zeta$ satisfy $|\eta-x|, |\zeta-x| < 1/n$ and are such that 
        \[ \begin{split} \| (\widetilde{G}_\scriptr(\eta)  - \widetilde{G}_\scriptr(\zeta)) v \| \geq & \| (\widetilde{G}_\scriptr(z) - \widetilde{G}_\scriptr(w)) u \| - \| (\widetilde{G}_\scriptr(\eta) - \widetilde{G}_\scriptr(z)) u \| \\ - & \| (\widetilde{G}_\scriptr(\zeta) - \widetilde{G}_\scriptr(w)) u \| - 2 \| \widetilde{G}_\scriptr \| \| u - v \| \\ > & \epsilon/3 \geq 1/k \end{split} \]
        by the fact that $\widetilde{G}_\scriptr$ is a Schur-class function. The reverse direction is immediate. Hence, we may write $N_G$ as a countable union and intersection of sets
        \[ \begin{split} \bigcup_{v \in D_{\text{ran}(P)}} \bigcup_{\beta = 2}^\infty \bigcup_{k = 1}^\infty \bigcap_{n = 1}^\infty \bigcup_{\eta,\zeta \in D_{\mathbb{D}}} \bigg\{(\scriptr, x) \in R \times \mathbb{T} \ | \ &\eta,\zeta \in \Gamma_\beta(x), |\eta - x|, |\zeta - x| < \frac{1}{n}, \\ &\|\widetilde{G}_\scriptr( \eta ) v - \widetilde{G}_\scriptr( \zeta ) v \| > \frac{1}{k} \bigg\} \text{.} \end{split}  \]
        Because $\scriptr \mapsto \widetilde{G}_\scriptr( \eta ) v - \widetilde{G}_\scriptr( \zeta ) v$ is measurable, so is $\scriptr \mapsto \|\widetilde{G}_\scriptr( \eta ) v - \widetilde{G}_\scriptr( \zeta ) v \|$ for fixed $\eta,\zeta \in \mathbb{D}$ and $v \in \text{ran}(P)$.  This implies that $\{\scriptr \in R \ | \ \|\widetilde{G}_\scriptr( \eta ) v - \widetilde{G}_\scriptr( \zeta ) v \| > \frac{1}{k} \}$ is in $\mathcal{R}$. Because the Stolz region is open, so is $\{x \in \mathbb{T} \ | \ \eta,\zeta \in \Gamma_\beta(x), |\eta - x|, |\zeta - x| < \frac{1}{n} \}$. Hence, $N_G$ is measurable in $\mu \times m$. It is clear that $\chi_{N_G}$ is integrable in $\mu \times m$ and that $\mu$ is ($\sigma$-)finite and thus 
        \[ 0 = \int_R m(N_G^\scriptr) \ d \mu(\scriptr) = \int_R \int_\mathbb{T} \chi_{N_G} \ dm(x) \ d \mu(\scriptr) = \int_\mathbb{T} \mu(N_G^x) \ d m(x) \]
        by Fubini-Tonelli, where $N_G^\scriptr := \{ x \in \mathbb{T} \ | \ (\scriptr, x) \in N_G \}, N_G^x := \{ \scriptr \in R \ | \ (\scriptr, x) \in N_G \}$ and $m(N_G^\scriptr) = 0$ by the existence of the non-tangential boundary limit a.e. on $\mathbb{T}$ for each $\scriptr \in R$. This implies that there exists some conull set $E \subseteq \mathbb{T}$ such that for each $x \in E$ we have that $\widetilde{G}_\scriptr$ has non-tangential limit at $x$ $\mu$-a.e. $\scriptr \in R$. Then the following exists for $(x,y)$ in $E \times E$
        \[ \begin{split} \varphi(x,y) :&= \lim_{z \rightarrow x, w \rightarrow y} \widetilde{\varphi}(z,w) = \lim_{z \rightarrow x, w \rightarrow y} \int_R \langle \widetilde{G}_\scriptr(z) | \widetilde{G}_\scriptr(w)\rangle_\text{HS} \ d \mu(\scriptr) \\ &= \int_R \lim_{z \rightarrow x, w \rightarrow y} \langle \widetilde{G}_\scriptr(z) | \widetilde{G}_\scriptr(w)  \rangle_\text{HS} \ d \mu(\scriptr) \end{split} \]
        by dominated convergence, where the limit in the integral is defined $\mu$-a.e. Moreover, $|\varphi(x,y)| \leq d \ \mu(R)$ on $E \times E$ and thus $\varphi$ is essentially bounded and equals the pointwise limit of measurable functions, and thus is measurable.

    \section{Proof of corollary \ref{cor:spec_decomp}}
    \label{ap:proof_of_spec_decomp}
    Let $[\widetilde{v}]$ in $H^\infty(\mathbb{D}, \mathcal{B}(\mathbb{C}, \mathcal{K}_d))$ be the vector from corollary \ref{cor:Stinespring_rep} and $[v]$ its boundary. Because the $v_i$'s form a minimal Stinespring representation, they are strongly linear independent and note that $v_i \in H^\infty(\mathbb{T})$. Then also $\eta_i(x) := v_i(x) \psi(x)$ are strongly linear independent. Indeed, suppose that they would not, then there exists a non-zero (square-summable) sequence $\{a_i\}$ such that
    \[ \sum_{i = 1}^d a_i v_i(x) \psi(x) = 0  \]
    a.e. on $\mathbb{T}$. Let $M \subset \mathbb{T}$ be the set on which $\psi \neq 0$, which has positive measure. Then $\sum_{i = 1}^d a_i v_i(x) = 0$ a.e. on $M$. The possibly infinite sum $\sum_{i = 1}^d a_i \widetilde{v}_i(z)$ converges and is analytic by Weierstrass' Theorem on uniform convergence. Hence, $\sum_{i = 1}^d a_i \widetilde{v}_i(z) = 0$ on $\mathbb{D}$ by the fact that its boundary function is zero on a set of positive measure, $M$. This implies that $\sum_{i = 1}^d a_i v_i(x) = 0$ a.e. on $\mathbb{T}$ which is a contradiction with strong linear independence. 
    
    The Gram matrix $G_{ij} := \langle \eta_i | \eta_j \rangle$ is trace-class because of
    \[ \operatorname{tr}[G] = \sum_i G_{ii} = \frac{1}{2\pi} \int_\mathbb{T} \sum_{i=1}^d |v_i(x)|^2 |\psi(x)|^2 dx \leq \| v \|_\infty^2 \|\psi \|^2  < \infty  \text{.} \]
    Then $G$ is normal and compact and henceforth permits a spectral decomposition $G = U \Lambda U^\ast$ with $\Lambda_{ij} = \lambda_i \delta_{ij}$ and $\lambda_i>0$ by strong linear independence. Define a.e. $u_i = M_{h_i} \psi$ with $h_i = (\lambda_i)^{-1/2} \sum_{k=1}^d U_{ik}^\top v_k$. The finite sum of the analytic extensions
    $\widetilde{h}_i^{(n)}(z) = (\lambda_i)^{-1/2} \sum_{k=1}^n U_{ik}^\top \widetilde{v}_k(z)$ is analytic and converges uniformly to $\widetilde{h}_i$ because
    \[ \| \widetilde{h}_i^{(n)} - \widetilde{h}_i \|_\infty \leq \lambda_i^{-1/2} \left(\sum_{k=n+1}^d | U_{ik}^\top |^2\right)^{1/2} \| \widetilde{v} \|_\infty \rightarrow 0 \]
    as $n \rightarrow d$. Thus $\widetilde{h}_i$ is analytic and the boundary limit $h_i$ is in $H^\infty(\mathbb{T})$. Then, observe that
    \[ \langle u_i | u_j \rangle = (\lambda_i \lambda_j)^{-1/2} \sum_{k, \ell=1}^d U_{ik}^\ast U_{j \ell}^\top \langle \eta_k | \eta_\ell \rangle = (\lambda_i \lambda_j)^{-1/2} \sum_{k, \ell, m, n=1}^d U_{ik}^\ast U_{\ell j} U_{km} \Lambda_{mn} U^\ast_{n \ell} = \delta_{ij} \]
    and
    \[ \begin{split} \varphi(x,y) \overline{\psi(x)} \psi(y) &= \sum_{i=1}^d \overline{v_i(x)}\overline{\psi(x)} v_i(y)\psi(y) = \sum_{i,j,k=1}^d U_{ij} (\lambda_j)^{1/2} \overline{u_j(x)} \overline{U_{ik}} (\lambda_k)^{1/2} u_k(y) \\ &= \sum_{i}^d \lambda_i \overline{u_i(x)} u_i(y) \end{split} \]

\section{Algorithms}

\label{ap:Algorithms}

\begin{algorithm}[H]

    \caption{Finding a basis of a reproducing kernel Hilbert space}

    \label{alg:RKHS_basis}

    \textbf{Input:} Finite reproducing kernel Hilbert space $\mathcal{H}$ with positive semidefinite kernel $K: X \times X \rightarrow \mathbb{C}$ and set $X$

    \textbf{Output:} Orthonormal basis of $\mathcal{H}$

    \begin{algorithmic}[1]

        \State Pick a spanning set of points $x := \{x_i\}_{i = 1}^n \subseteq X$

        \State Calculate the Gram matrix $G_{ij} \gets K(x_i, x_j)$

        \State Calculate row-echelon form $(R, \mathcal{I}) \gets \mathrm{RREF}(G)$ with pivots $\mathcal{I} \subseteq \{1, \dots, n \}$

        \State Set $G' \gets G_{\mathcal{I} \mathcal{I}}$

        \State Calculate $L \gets \mathrm{Cholesky}(G')$

        \State Calculate $e \gets \overline{L^{-1}} K(\cdot, x_{\mathcal{I}})$

        \State \Return $(e, L, x, \mathcal{I})$

    \end{algorithmic}

\end{algorithm}

\begin{corollary}[Algorithm \ref{alg:RKHS_basis}]

    $G'$ is a positive definite matrix and $\{e_i\}_{i = 1}^r$ is an orthonormal basis of $\mathcal{H}$.

    \begin{proof}

        Because $G$ is Hermitian, we know that $G$ has rank $r$ if and only if $G$ has an $r \times r$ principal submatrix which is nonsingular, and no larger principal submatrix which is nonsingular (Theorem 5.19 in \cite{Perlis1992}). The columns $G_{\cdot \mathcal{I}}$ are a linearly independent basis of $\text{col}(G)$ and $r = |\mathcal{I}| = \operatorname{rank}(G)$. Then, by the principal submatrix rank property \cite{Berman2005} the matrix $G' = G_{\mathcal{I} \mathcal{I}}$ will be a positive definite, $r \times r$ principal submatrix of $G$.

        Secondly, observe that for all $i,j = 1, \dots, d$ we have

        \[
        \begin{split}
            \langle e_i | e_j \rangle_{\mathcal{H}}
            &=
            \sum_{k,l = 1}^d
            L_{ik}^{-1}
            \langle
                K(\cdot, x_{\mathcal{I}_k})
                |
                K(\cdot, x_{\mathcal{I}_l})
            \rangle_{\mathcal{H}}
            \overline{L_{jl}^{-1}}
            \\
            &=
            \sum_{k,l = 1}^d
            L_{ik}^{-1}
            G'_{kl}
            (L^\ast)_{lj}^{-1}
            \\
            &=
            \sum_{k,l,m = 1}^d
            L_{ik}^{-1}
            L_{km}
            L^\ast_{ml}
            (L^\ast)_{lj}^{-1}
            =
            \delta_{ij}
        \end{split}
        \]
        completing the proof.

    \end{proof}

\end{corollary}

\begin{algorithm}[H]

    \caption{Calculation of the complementary factor of a left inner column-valued function}

    \label{alg:complementary_factor}

    \textbf{Input:} Rational, left inner function $\widetilde{\Theta} \in H^\infty(\mathbb{D}, \mathcal{B}(\mathbb{C}, \mathbb{C}^d))$ with finite $d$ 

    \textbf{Output:} Rational, two-sided inner completion, $[\widetilde{\Theta}, \widetilde{\Theta}^c]$, and a minimal, conservative realization, $\widehat{U}^\ast$

    \begin{algorithmic}[1]

        \State Compute an orthonormal basis $(e, L, x, \mathcal{I}) \gets \mathrm{Alg_1}(\mathcal{H}(\widetilde{\Theta}^\sharp), K_{\widetilde{\Theta}^\sharp}, \mathbb{D})$ such that $0 \not \in x$

        \State Compute the canonical realization $U$ in terms of basis $\{e_i\}$ in the sense that

        \begin{enumerate}

            \item $A^{m n} \gets \sum_{\ell = 1}^r \overline{L^{-1}_{m \ell}} (e_n(x_{\mathcal{I}_\ell}) - e_n(0)) /x_{\mathcal{I}_\ell}$

            \item $B^{m n} \gets \sum_{\ell = 1}^r \overline{L^{-1}_{m \ell}} (\widetilde{\Theta}^\sharp_{1n}(x_{\mathcal{I}_\ell}) - \widetilde{\Theta}^\sharp_{1n}(0)) /x_{\mathcal{I}_\ell}$

            \item $C^{1 n} \gets e_n(0)$

            \item $D^{m n} \gets \widetilde{\Theta}^\sharp_{mn}(0)$

        \end{enumerate}

        \State Compute orthonormal basis matrix $W \gets \mathrm{orthonormal}(\ker(U))$

        \State Set $\widehat{U} \gets \begin{pmatrix} {U} \\ W^\ast \end{pmatrix}$

        \State Calculate the characteristic function $\theta_{\widehat{U}^\ast}$

        \State \Return $(\theta_{\widehat{U}^\ast}, \widehat{U}^\ast)$

    \end{algorithmic}

\end{algorithm}

\begin{algorithm}[H]

    \caption{Finding the zeroes and projectors of a rational two-sided inner function}

    \label{alg:blaschke_potapov}

    \textbf{Input:} Rational two-sided inner function $B : \mathbb{D} \rightarrow \mathcal{B}(\mathbb{C}^d)$ with $\deg B < \infty$

    \textbf{Output:} Blaschke-Potapov factors $\{B_k\}_{k=1,\ldots,N}$ such that $B(z) = U_0 \overset{\curvearrowright}{\prod}_{k=N}^1 B_k(z)$ for some constant unitary $U_0$

    \begin{algorithmic}[1]

        \State Initialize residual $R(z) \gets B(z)$

        \State Initialize known Blaschke-Potapov product $B_{\mathrm{known}}(z) \gets I$

        \While{$\det R$ has a zero in $\mathbb{D}$}

            \State Pick $a_k \in \mathbb{D}$ such that $\det R(a_k) = 0$

            \State Compute orthonormal basis $\{v_{ki}\}_{i=1}^r \gets \mathrm{orthonormal}(\ker(R(a_k)))$

            \State Build projector $P_k \gets \sum_{i=1}^r v_{ki} v_{ki}^{\dagger}$

            \State Build elementary factor $B_k(z) \gets I + (b_{a_k}(z) - 1) P_k$

            \State Update known product $B_{\mathrm{known}}(z) \gets B_k(z)B_{\mathrm{known}}(z)$

            \State Update residual $R(z) \gets R(z)B_k(z)^{-1}$

        \EndWhile

        \State Pick $z_0 \in \mathbb{D}$ such that $\det B_{\mathrm{known}}(z_0) \neq 0$

        \State Set $U_0 \gets B(z_0)B_{\mathrm{known}}(z_0)^{-1}$

        \State \Return $(U_0, \{B_k\}_{k=1,\ldots,N})$

    \end{algorithmic}

\end{algorithm}

\begin{corollary}[Algorithm \ref{alg:blaschke_potapov}]

    For a finite Blaschke-Potapov product of $N$ terms, the algorithm completes and produces $(U_0, \{B_k\}_{k=1,\ldots,N})$ such that
    \[
        U_0 \overset{\curvearrowright}{\prod}_{k=N}^1 B_k(z) = B(z)
    \]
    for all $z \in \mathbb{D}$.

    \begin{proof}

        Consider first that each rational two-sided inner function, $R$, can be expressed as some Blaschke-Potapov product
        \[
            R(z)
            =
            U
            (I + (b_1(z) - 1) P_1)
            \cdots
            (I + (b_M(z) - 1) P_M)
        \]
        for some orthoprojectors $P_i$, unitary $U$ and
        \[
            b_{a}(z)
            = \begin{cases}
            \frac{|a|}{a}
            \frac{z-a}{1-\overline{a} z} 
            \quad
            a \in \mathbb{D} \setminus \{ 0 \} \\ z \qquad \qquad a=0 \end{cases}
        \text{.} \]
        Moreover, $R(z)$ satisfies
        \[
            \det R(z)
            =
            (\det U)\,
            b_{a_1}(z)^{\operatorname{rank} P_1}
            \cdots
            b_{a_M}(z)^{\operatorname{rank} P_M}
            =
            \frac{p(z)}{q(z)}
        \]
        where $p$ and $q$ are coprime polynomials with
        $\deg q \leq \deg p = \sum_{i=1}^M \operatorname{rank} P_i$.
        Now let $a \in \mathbb{D}$ be such that $\det R(a) = 0$ and define
        $P_{\ker R(a)}$ as the orthoprojector on the kernel of $R_a$.
        Then
        \[
            B_a(z)
            =
            I + (b_a(z) - 1)P_{\ker R(a)}
        \]
        is a Blaschke-Potapov term that is invertible away from $z = a$ such that we may write
        \[
            \widetilde{R}(z) = R(z)B_a(z)^{-1}.
        \]
        Update $\widetilde{B}_\text{known}(z) = B_a(z) B_\text{known}(z)$ and observe that
        \[ \widetilde{R}(z) \widetilde{B}_\text{known}(z) = R(z) B_a(z)^{-1} B_a(z) B_\text{known}(z) = R(z) B_\text{known}(z) = B(z) \text{.} \]
        Expand $R(z) = R(a) + (z-a) H(z)$ for some sufficiently small neighborhood around $z = a$ and some analytic matrix-valued function $H(z)$. We find that
        \[
            \widetilde{R}(z)
            =
            R(z)P_{\ker R(a)}^\perp
            +
            \begin{cases}
            \frac{a}{|a|} (1 - \overline{a} z)
            H(z) P_{\ker R(a)} \quad a \neq 0 \\
            (1 - \overline{a} z)
            H(z) P_{\ker R(a)} \qquad a = 0
            \end{cases}
        \]
        which is analytic around $z = a$. Since moreover $B_a(x)$ is unitary for $x \in \mathbb{T}$, $\widetilde{R}(z)$ is again a rational two-sided inner function that satisfies
        \[
            \det \widetilde{R}(z)
            =
            \frac{\det R(z)}
            {b_a(z)^{\dim \ker R(a)}}
        \]
        such that the degree decreases. Repeating this argument will give a rational two-sided inner function $R_0(z)$ such that for some $c \in \mathbb{C}$ we have $\det R_0(z) = c$. Now given the fact that $c \, q_0(z) = p_0(z)$ for some coprime polynomials $p_0, q_0$, we find that $p_0$ and $q_0$ are both constant, implying that $R_0(z) = U_0$ for some unitary $U_0$ and all $z \in \mathbb{D}$. And observe that because in each round $\widetilde{R}(z) \widetilde{B}_\text{known}(z) = B(z)$, we have $B(z) = U_0 \widetilde{B}_\text{known}(z)$ finally. 
    \end{proof}

\end{corollary}

\end{appendix}
\end{appendices}

\bibliography{references}

@book{Nikolski2002,
  author    = {Nikolai K. Nikolski},
  title     = {Operators, Functions, and Systems: An Easy Reading. Volume 1: {Hardy}, {Hankel}, and {Toeplitz}},
  series    = {Mathematical Surveys and Monographs},
  volume    = {92},
  publisher = {American Mathematical Society},
  address   = {Providence, RI},
  year      = {2002}
}

@book{Nikolski2002_2,
  author    = {Nikolai K. Nikolski},
  title     = {Operators, Functions, and Systems: An Easy Reading. Volume 2: Model Operators and Systems},
  series    = {Mathematical Surveys and Monographs},
  volume    = {93},
  publisher = {American Mathematical Society},
  address   = {Providence, RI},
  year      = {2002}
}

@book{Watrous2018,
  author    = {John Watrous},
  title     = {The Theory of Quantum Information},
  publisher = {Cambridge University Press},
  address   = {Cambridge},
  year      = {2018},
  doi       = {10.1017/9781316848142}
}

@book{Reed1980,
  author    = {Michael Reed and Barry Simon},
  title     = {Methods of Modern Mathematical Physics, Volume {I}: Functional Analysis},
  edition   = {Revised and enlarged},
  publisher = {Academic Press},
  address   = {San Diego},
  year      = {1980}
}

@book{Holevo2012,
  author    = {Alexander S. Holevo},
  title     = {Quantum Systems, Channels, Information: A Mathematical Introduction},
  series    = {De Gruyter Studies in Mathematical Physics},
  volume    = {16},
  publisher = {De Gruyter},
  address   = {Berlin},
  year      = {2012},
  doi       = {10.1515/9783110273403}
}

@book{Nielsen2010,
  author    = {Michael A. Nielsen and Isaac L. Chuang},
  title     = {Quantum Computation and Quantum Information},
  edition   = {10th anniversary},
  publisher = {Cambridge University Press},
  address   = {Cambridge},
  year      = {2010},
  doi       = {10.1017/CBO9780511976667}
}

@article{Duquet2022,
  author  = {Charles Duquet},
  title   = {Dilation properties of measurable {Schur} multipliers and {Fourier} multipliers},
  journal = {Positivity},
  volume  = {26},
  number  = {4},
  pages   = {69},
  year    = {2022},
  doi     = {10.1007/s11117-022-00933-x}
}

@article{Steen2012,
  author  = {Naomi M. Steen and Ivan G. Todorov and Lyudmila Turowska},
  title   = {Local operator multipliers and positivity},
  journal = {J. Funct. Anal.},
  volume  = {267},
  number  = {1},
  pages   = {80--111},
  year    = {2014},
  doi     = {10.1016/j.jfa.2014.04.007}
}

@book{Steinwart2008,
  author    = {Ingo Steinwart and Andreas Christmann},
  title     = {Support Vector Machines},
  series    = {Information Science and Statistics},
  publisher = {Springer},
  address   = {New York},
  year      = {2008},
  doi       = {10.1007/978-0-387-77242-4}
}

@article{Birman2003,
  author  = {Mikhail Sh. Birman and Michael Solomyak},
  title   = {Double operator integrals in a {Hilbert} space},
  journal = {Integral Equ. Oper. Theory},
  volume  = {47},
  number  = {2},
  pages   = {131--168},
  year    = {2003},
  doi     = {10.1007/s00020-003-1157-8}
}

@article{Curto2021,
   author = {Raúl E. Curto and In Sung Hwang and Woo Young Lee},
   doi = {10.1016/J.JFA.2020.108884},
   issn = {10960783},
   issue = {6},
   journal = {Journal of Functional Analysis},
   month = {3},
   publisher = {Academic Press Inc.},
   title = {The {Beurling-Lax-Halmos Theorem} for infinite multiplicity},
   volume = {280},
   url = {https://www.researchgate.net/publication/347781240_The_Beurling-Lax-Halmos_theorem_for_infinite_multiplicity},
   year = {2021}
}

@book{Paulsen2016,
  author    = {Vern I. Paulsen and Mrinal Raghupathi},
  title     = {An Introduction to the Theory of Reproducing Kernel {Hilbert} Spaces},
  publisher = {Cambridge University Press},
  address   = {Cambridge},
  year      = {2016},
  doi       = {10.1017/CBO9781316219232}
}

@book{Perlis1992,
  author    = {Sam Perlis},
  title     = {Theory of Matrices},
  publisher = {Dover Publications},
  address   = {New York},
  year      = {1991},
  note      = {Dover reprint of the 1952 Addison--Wesley edition}
}

@article{Berman2005,
  author  = {Abraham Berman and Thomas J. Laffey},
  title   = {Similarity classes and principal submatrices},
  journal = {Linear Algebra Appl.},
  volume  = {401},
  pages   = {341--351},
  year    = {2005},
  doi     = {10.1016/j.laa.2004.09.013}
}

@incollection{Ball2006,
  author    = {Joseph A. Ball and Vladimir Bolotnikov and Quanlei Fang},
  title     = {{Schur}-class multipliers on the {Fock} space: {de Branges--Rovnyak} reproducing kernel spaces and transfer-function realizations},
  booktitle = {Operator Theory, Structured Matrices, and Dilations: {Tiberiu Constantinescu} Memorial Volume},
  editor    = {Mihaly Bakonyi and Aurelian Gheondea and Mihai Putinar and James Rovnyak},
  pages     = {85--114},
  publisher = {Theta Foundation},
  address   = {Bucharest},
  year      = {2007}
}

@incollection{Fritzsche2008,
  author    = {Bernd Fritzsche and Victor Katsnelson and Bernd Kirstein},
  title     = {The {Schur} algorithm in terms of system realizations},
  booktitle = {Characteristic Functions, Scattering Functions and Transfer Functions: The {Moshe Livsic} Memorial Volume},
  editor    = {Daniel Alpay and Victor Vinnikov},
  series    = {Operator Theory: Advances and Applications},
  volume    = {197},
  pages     = {181--250},
  publisher = {Birkh{\"a}user},
  address   = {Basel},
  year      = {2010},
  doi       = {10.1007/978-3-0346-0183-2_9}
}

@incollection{Frazho2018,
  author    = {A. E. Frazho and A. C. M. Ran},
  title     = {A note on inner--outer factorization of wide matrix-valued functions},
  booktitle = {Operator Theory, Analysis and the State Space Approach: In Honor of {Rien Kaashoek}},
  editor    = {Harm Bart and Sanne ter Horst and A. C. M. Ran and Hugo J. Woerdeman},
  series    = {Operator Theory: Advances and Applications},
  volume    = {271},
  pages     = {201--214},
  publisher = {Birkh{\"a}user},
  address   = {Cham},
  year      = {2018},
  doi       = {10.1007/978-3-030-04269-1_8}
}

@phdthesis{Ende2020,
  author  = {Frederik vom Ende},
  title   = {Reachability in Controlled {Markovian} Quantum Systems: An Operator-Theoretic Approach},
  school  = {Technical University of Munich},
  address = {Munich},
  year    = {2020},
  doi     = {10.14459/2020md1559809}
}

@article{Deutsch2010,
  author  = {Frank Deutsch and Hein Hundal},
  title   = {Slow convergence of sequences of linear operators {I}: almost arbitrarily slow convergence},
  journal = {J. Approx. Theory},
  volume  = {162},
  number  = {9},
  pages   = {1701--1716},
  year    = {2010},
  doi     = {10.1016/j.jat.2010.05.001}
}

@book{Paulsen2003,
  author    = {Vern I. Paulsen},
  title     = {Completely Bounded Maps and Operator Algebras},
  series    = {Cambridge Studies in Advanced Mathematics},
  volume    = {78},
  publisher = {Cambridge University Press},
  address   = {Cambridge},
  year      = {2003},
  doi       = {10.1017/CBO9780511546631}
}

@misc{Wolf2023,
  author       = {Michael M. Wolf},
  title        = {Mathematical Introduction to Quantum Information Processing},
  howpublished = {Lecture notes, Technical University of Munich},
  year         = {2023},
  month        = mar,
  url          = {https://mediatum.ub.tum.de/1706981},
  note         = {Growing lecture notes}
}

@article{Aleksandrov2025,
  author  = {Aleksei B. Aleksandrov and Vladimir V. Peller},
  title   = {Analytic {Schur} multipliers},
  journal = {Proc. Am. Math. Soc.},
  volume  = {154},
  number  = {3},
  pages   = {1207--1216},
  year    = {2026},
  doi     = {10.1090/proc/17531}
}

@incollection{Ball2014,
  author    = {Joseph A. Ball and Vladimir Bolotnikov},
  title     = {{de Branges--Rovnyak} spaces: basics and theory},
  booktitle = {Operator Theory},
  editor    = {Daniel Alpay},
  pages     = {631--679},
  publisher = {Springer},
  address   = {Basel},
  year      = {2015},
  doi       = {10.1007/978-3-0348-0667-1_6}
}

@book{Rabus2007,
  author    = {Dominik G. Rabus},
  title     = {Integrated Ring Resonators: The Compendium},
  series    = {Springer Series in Optical Sciences},
  volume    = {127},
  publisher = {Springer},
  address   = {Berlin, Heidelberg},
  year      = {2007},
  doi       = {10.1007/978-3-540-68788-7}
}

@incollection{Katsnelson2007,
  author    = {Victor E. Katsnelson and Bernd Kirstein},
  title     = {On the theory of matrix-valued functions belonging to the {Smirnov} class},
  booktitle = {Topics in Interpolation Theory},
  editor    = {Harry Dym and Victor Katsnelson and Bernd Fritzsche and Bernd Kirstein},
  series    = {Operator Theory: Advances and Applications},
  volume    = {95},
  pages     = {299--350},
  publisher = {Birkh{\"a}user},
  address   = {Basel},
  year      = {1997},
  doi       = {10.1007/978-3-0348-8944-5_14}
}

@book{Staffans2005,
  author    = {Olof J. Staffans},
  title     = {Well-Posed Linear Systems},
  series    = {Encyclopedia of Mathematics and its Applications},
  volume    = {103},
  publisher = {Cambridge University Press},
  address   = {Cambridge},
  year      = {2005},
  isbn      = {9780521825849}
}

@article{Lilleberg2019,
   author = {Lassi Lilleberg},
   doi = {10.1007/s11785-019-00930-1},
   journal = {Complex Analysis and Operator Theory},
   pages = {3767-3793},
   title = {Passive Discrete-Time Systems with a {Pontryagin} State Space},
   volume = {13},
   url = {https://doi.org/10.1007/s11785-019-00930-1},
   year = {2019}
}

\end{document}